\documentclass[]{aa} 

\pdfoutput=1

\usepackage{lscape}
\usepackage{longtable}
\usepackage{graphicx}
\usepackage{times}
\usepackage{graphicx}
\usepackage{xspace}
\usepackage{epsfig}
\usepackage{natbib}
\usepackage{rotating}
\usepackage{dcolumn}
\usepackage{xcolor}
\usepackage{txfonts}

\usepackage{hyperref}
\usepackage{cleveref}

\begin{document}


   \title{A search for pre- and proto-brown dwarfs in the dark cloud Barnard~30 with ALMA}


\author{
          N. Hu\'elamo
          \inst{1}
          \and
          I. de Gregorio-Monsalvo\inst{2,3}
          \and
         A. Palau
          \inst{4}
          \and
          D. Barrado
          \inst{1}
          \and
          A. Bayo
          \inst{5,6}
          \and
          M.T. Ruiz
          \inst{7}
          \and
          L. Zapata
          \inst{4}  
          \and
         H. Bouy
         \inst{1}
          \and
          O. Morata
          \inst{8}
         \and
         M. Morales-Calder\'on
         \inst{1}
          \and
          C. Eiroa
          \inst{9}
          \and
          F. M\'enard
          \inst{10}
}
   \offprints{N. Hu\'elamo}

\institute{
  Dpto. Astrof\'{\i}sica, Centro de Astrobiolog\'{\i}a (INTA-CSIC);  ESAC campus, Camino bajo del Castillo s/n, Urb. Villafranca del Castillo, E-28692 Villanueva de la Ca\~nada, Spain \\  \email{nhuelamo@cab.inta-csic.es}
         \and
             European Southern Observatory, Alonso de C\'ordova 3107, Vitacura, Santiago, Chile.
             \and
            Joint ALMA office, Alonso de C\'ordova 3107, Vitacura, Casilla 19001, Santiago 19, Chile 
         \and
             Instituto de Radioastronom\'{\i}a y Astrof\'{\i}sica, Universidad Nacional Aut\'onoma de M\'exico, P.O. Box 3-72, 58090 Morelia, Michoac\'an, M\'exico
         \and
             Departamento de F\'{\i}sica y Astronom\'{\i}a, Facultad de Ciencias, Universidad
             de Valpara\'{\i}so, Av. Gran Breta\~na 1111, 5030 Casilla, Valpara\'{\i}so,  Chile
             \and 
             ICM nucleus on protoplanetary disks, Universidad de Valpara\'{\i}so, Av. Gran Breta\~na 1111, Valpara\'{\i}so, Chile
         \and
         Dpto. de Astronom\'{\i}a, Universidad de Chile, Camino del Observatorio 1515, Santiago, Chile
         \and
           Institute of Astronomy and Astrophysics, Academia Sinica, 11F of Astronomy-Mathematics Building, AS/NTU. No.1, Sec. 4, Roosevelt Rd, Taipei 10617, Taiwan          \and
            Unidad Asociada Astro-UAM, UAM, Unidad Asociada CSIC, Universidad Aut\'onoma de Madrid, Ctra. Colmenar km.~15, Madrid E-28049, Spain
             \and
            Institut de Plan\'etologie et d'Astrophysique de Grenoble (UMR 5274),  BP 53, F-38041 Grenoble,  C\'edex 9, France
}

   \date{Received ; accepted }

 
  \abstract
   {The origin of brown dwarfs is still under debate. While some models predict a star-like formation scenario, others invoke a substellar mass 
   embryo ejection, a stellar disk fragmentation, or the photo-evaporation of an external core due to the presence of massive stars.
   }
   {The aim of our work is to characterize the youngest and lowest mass population of the dark cloud Barnard 30, a region within the Lambda Orionis star-forming region. 
   In particular,  we aim to identify proto-brown dwarfs and study the mechanism of their formation.}
   {We obtained  ALMA  continuum observations at 880\,$\mu$m of 30 sub-mm cores previously identified with APEX/LABOCA at 870\,$\mu$m.  We have complemented part of the ALMA data with sub-mm APEX/SABOCA observations at 350\,$\mu$m, and with multi-wavelength ancillary observations from the optical to the far-infrared (e.g., Spitzer, CAHA/O2000, WISE, INT/WFC).
   }
   {We report the detection of five (out of 30) spatially unresolved sources with ALMA, with estimated masses between 0.9 and 67\,M$_{\rm Jup}$. From these five sources, only two show gas emission.   The analysis of multi-wavelength photometry from these two objects, namely B30-LB14 and B30-LB19,  is consistent with one Class~II- and one Class~I low-mass stellar object, respectively. The gas emission is consistent with a rotating disk in the case of B30-LB14, and with an oblate rotating envelope with infall signatures in the case of LB19.  The remaining three ALMA detections do not have infrared counterparts and can be classified as either deeply embedded objects or as starless cores if B30 members. In the former case, two of them (LB08 and LB31) show internal luminosity upper limits consistent with Very Low Luminosity objects, while we do not have enough information for LB10. In the starless core scenario, and taking into account the estimated masses from ALMA and the  APEX/LABOCA cores, we estimate final masses for the central objects in the substellar domain, so they could be classified as pre-BD core candidates. According to the turbulent fragmentation scenario, the three ALMA pre-BD core candidates should be gravitationally stable based on APEX/LABOCA data. However, this result is not consistent with the presence of compact sources inside the cores, indicative of on-going collapse. As an alternative scenario we propose that these cores 
   could be the result of on-going gravitational contraction. Indeed, we have verified that their estimated masses are consistent with the ones expected within an ALMA beam for a $r^{-2}$ density profile, which is typical for a collapsing core.  }
   {ALMA observations have allowed us to detect very low-mass compact sources within three  APEX/LABOCA cores. Future observations will help us to unveil their true nature.}


   \maketitle
%

%
\section{Introduction \label{intro}}
%

Brown dwarfs (BDs) are the bridge between low-mass stars and Jupiter-like planets.
Although abundant in star forming regions \citep[e.g.,][]{White2003,Barrado2004.3,Caballero2007,Bouy2007,Bayo2011.1,Alves2012,Scholz2013, Muzic2015}, and the field \citep[e.g.,][]{Delfosse1997,Delorme2008,Mainzer2011}, their formation mechanism is still under debate.
In low-mass star-forming regions,  which typically form objects in groups or loose associations, the main brown dwarf formation mechanisms
 are turbulent fragmentation \cite[][]{Padoan2004,Hennebelle2008}
 and ejection from multiple protostellar systems and/or fragmented disks
 \cite[][]{Reipurth2001,Bate2002,Matzner2005,Whitworth2006}. 
However, in the surroundings of high-mass stars, where objects typically form in a more tightly packed manner,
 there are additional mechanisms, namely photo-evaporation of cores near massive stars 
\cite[e.g.,][] {Hester1996,Whitworth2004} that  have already been observed in young clusters \citep[see e.g.,][]{Bouy2009,Hodapp2009}, 
and gravitational fragmentation of dense filaments formed in a nascent cluster \cite[e.g.,][]{Bonnell2008,Bate2012}.

The different BD formation scenarios can be tested by studying the earliest phases in the BD formation, when BDs are still embedded in the parental cloud, the so-called proto-BDs. The study of the most embedded and youngest BDs  requires observations in the mm/sub-mm regime, where they emit the bulk of their energy as they are dominated by cold envelopes \citep[e.g.,][]{Bourke2006,Lee2013,Palau2014}. 

Several studies have tried to unveil the youngest populations of BDs. The so-called Very Low Luminosty Objects 
\citep[VeLLOs, e.g.,][]{Young2004, diFrancesco2007,Dunham2008,Lee2013}, which are characterized by internal luminosities $\leq$0.1\,L$_{\odot}$, were identified as potential proto-BDs. The so-called pre-brown dwarf (pre-BD) cores are also potential proto-BDs. They are cores with no infrared/optical counterparts that could be gravitationally bound but have very low masses (near substellar) and thus they will not be able to form a stellar object even if accreting all the reservoir of mass within the core. Good examples of pre-BD cores can be found in e.g. \citet{Palau2012} or \citet{Andre2012}.
Up to now,  several objects have been identified as VeLLOs whose properties are consistent with proto-BDs 
\citep[see e.g.,][]{Young2004,Bourke2006,Lee2009,Barrado2009,Palau2012, Palau2014,Morata2015}. 
However, the samples of proto-BD candidates are still very small.

\begin{figure*}[ht!]
\includegraphics[width=1.0\textwidth,scale=0.50]{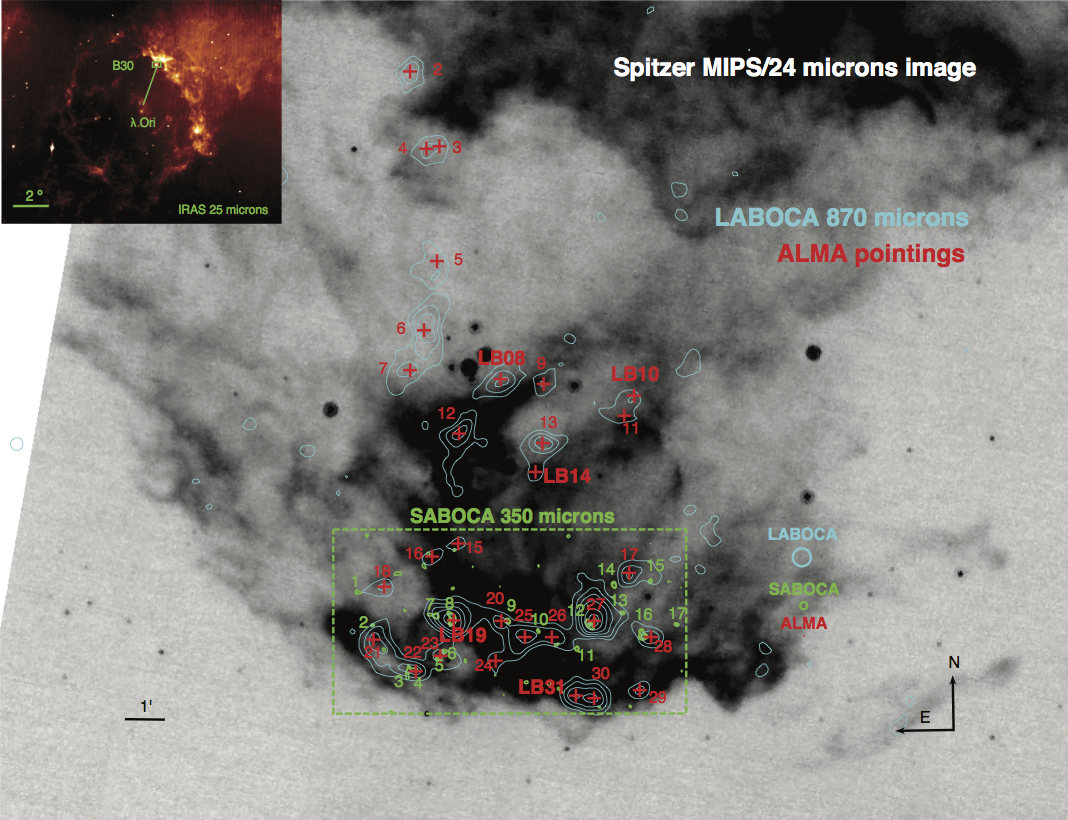}
\caption{Spitzer 24\,$\mu$m image of the B30 cloud. The displayed field of view is $\sim$ 23'$\times$ 18'.  The cyan contours represent the APEX/LABOCA map at  870\,$\mu$m discussed in BGH16 (we show 3, 5, 7, 9, 12 and 15-$\sigma$ levels).  The red crosses show the ALMA pointing coordinates, and correspond to the pixels with maximum intensity of the  LABOCA cores. The cores are numbered from 2 to 31 following BGH16. The five sources detected with ALMA are  marked with red  labels. The green contours (2.5, 3, 4 and 5-$\sigma$ levels) represent the APEX/SABOCA data, and the green box the total field of view of the APEX/SABOCA map. We have numbered the 17~APEX/SABOCA sources that are detected above the 3-$\sigma$ level. The beam sizes of each dataset are included on the right part of the image. Finally, the image
at the top left corner shows the whole LORI region as seen by IRAS at 25 $\mu$m, and the green box represents the B30 area studied with APEX/LABOCA and ALMA.}\label{fig:wideALMA}
\end{figure*}

In order to shed light on the BD formation process, in 2007 we started a project to study the low mass population of the dark cloud Barnard 30 (B30). B30 is located at the rim of the Lambda Orionis Star-Forming Region \citep[LOSFR,][]{Murdin77.1},  a complex structure  at a distance of $\sim$400\,pc \citep{Perryman1997, Dolan1999.1}, which gives shape to the head of Orion. The central part of the region is occupied by Collinder\,69,  a $\sim$5\,Myr cluster \citep[e.g.,][]{Dolan1999.1, Barrado2004.1,Barrado2007.1, Bayo2011.1} that harbours the O8~III star $\lambda$ Ori, while B30 is located $\sim$2.5 degrees North West from this massive star (see Figure~\ref{fig:wideALMA}). The B30 stellar association was first identified by \citet{Duerr82} and confirmed by \citet{Gomez98}.
Later, the stellar population was studied by \citet[][]{Dolan1999.1,Dolan2001.1,Dolan2002.1} who built a  
census of  pre-main sequence (PMS) stars in the region. Using mid-IR data from the IRAC instrument onboard the Spitzer telescope,  together with deep optical and near-IR data, \cite{Morales2008.1} was able to  characterize the young low-mass population of the cloud, and to identify new candidate members with masses between 1.1 and 0.02\,M$_{\odot}$ reaching  the substellar mass regime. Finally, \citet{Bayo2009.1} built the census of $\sim$70 spectroscopically confirmed members (including 15 new members), derived large accretion rates and accretion fractions, and  calculated an optimal age estimation of between one and three\,Myr given the large dispersion  in the HR diagram.
  
 To further unveil the youngest population of B30, \citet[][submitted; BGH16, hereafter]{Barrado2016} analyzed APEX/LABOCA 870\,$\mu$m deep observations (average $rms$ of 7\,mJy) of a region of  $\sim$0.5$^{\circ}$$\times$0.5$^{\circ}$ within the cloud, and  reported the detection of 34 sub-mm sources. They found IR counterparts for 15 of them within 5" from  the LABOCA peak intensity coordinates, with nine being good substellar candidates. Twelve objects did not show IR counterparts within 5", but showed several counterparts between 5 and 27 arcsec (the LABOCA image resolution) that might be responsible for the sub-mm emission. Nine of them (out of twelve) might also be substellar. Unfortunately,  the low spatial resolution of the LABOCA map ($\sim$27 arcsec) did not allow them to assign unambiguous infrared counterparts to most of the  sub-mm sources. Finally, they also reported five potential starless cores, that might be pre-BD core candidates according to their estimated masses. As a result of their work, BGH16 built a census of proto- and pre-BD candidates in B30.  
 We note that, recently, \citet{Liu2016} have also  presented  a study of  cold cores in the B30 region, reporting the discovery 
of a Class~0 protostar and a proto-BD candidate, but  in a different region of the cloud (at $\sim$27' North West from the APEX/LABOCA map obtained by BGH16).  

 In this work, we have further investigated the nature of the LABOCA sources identified by BGH16 through high angular-resolution and high-sensitivity observations of B30 with the Atacama Large Millimeter Array (ALMA). The superb spatial resolution, great sensitivity, and positional accuracy of ALMA can help us to unambiguously assign the correct counterparts to the sub-mm sources, and to study their spectral energy distributions and dust content. We have complemented part of the ALMA observations with sub-mm data obtained with APEX/SABOCA at 350\,$\mu$m. The observations are described in Section~\ref{sec:observations}, 
 while the data analysis is included in Section~\ref{sec:analysis}. The main results and conclusions are presented in Sections~4 and~5.
 

\section{Observations\label{sec:observations}}

\subsection{ALMA observations}

\begin{figure*} [t!]
\includegraphics[width=1.0\textwidth,scale=0.50]{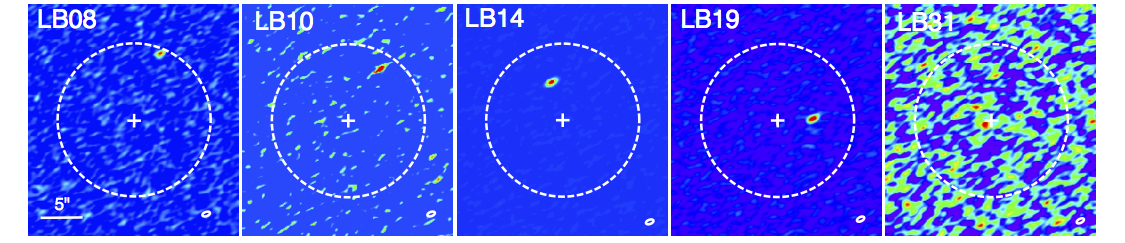}
\caption{\label{fig:ALMAdet} The five sources detected at 880\,$\mu$m with ALMA in B30. 
North is up and east to the left. The ALMA beam size is displayed at  the bottom right corner of each image.
The white dashed circles represent the ALMA primary beam (18" diameter), and are centered at the pointing coordinates (phase centers) for each source.}
\end{figure*}

\begin{table*}\caption{ALMA 880\,$\mu$m detections in the B30 region}\label{table:almadets}
{\tiny
\begin{tabular}{lllllllllll}\hline
 \multicolumn{2}{l}{ALMA detection coordinates} 
 & Separation from
 &  LABOCA  & S$_{\nu}$& $rms$ 
 & Mass  
 & Mass  
 & Missing  
 &  $A_{\rm v}$ $^{\rm d}$
 & 
 Tentative
 \\ 
 RA(J2000)  
 & DEC(J2000) 
 & phase center
 & source$^{\rm a}$     
 & &
 & (ALMA)$^{**}$
 & (LABOCA)$^{\rm b,**}$
 & flux$^{\rm c}$
 & 
 & nature
 \\
 $[\mathrm{h\,m\,s}]$
 &  [$^\circ$ \,\arcmin \, \arcsec]     
 & [arcsec]    
 &   
 & [mJy]    
 & [mJy/beam]
 &  [M$_{\rm Jup}$] 
 &  [M$_{\rm Jup}$] 
 & [\%] 
 & [mag ] 
 & 
 \\ \hline
               05:31:22.97  & +12:11:34.7       & 8.7 & B30-LB08                &  5.70   & 0.26& 9       & 106  & 87 & 2.5  & VeLLO/pre-BD\\ 
               05:31:09.29  & +12:11:08.8       & 6.8 & B30-LB10                & 2.20    & 0.23& 3       &  46      & 93 & 1.2 & pre-BD\\
               05:31:19.46  & +12:09:15.1       & 4.5 & B30-LB14$^*$    & 44.0    & 0.25& 67         & 51       & 0   & 2.0 & YSO\\
               05:31:27.81  & +12:05:30.9       & 4.2 & B30-LB19$^*$    & 10.7            & 0.20& 16         & 182  & 86 & 2.8  & YSO\\
               05:31:15.32  & +12:03:38.2       & 0.6 & B30-LB31                & 0.60            & 0.13 & 0.9    &  82      & 99   & 2.4 & VeLLO/pre-BD\\
               \hline
              \end{tabular}
              }
\tablefoot{ 
\tablefoottext{\rm a}{LABOCA designation number (see BGH16); }  
\tablefoottext{\rm b}{From BGH16; }
\tablefoottext{\rm c}{ Percentage of missing flux between ALMA and LABOCA data; }
\tablefoottext{\rm d}{From the 2MASS extinction map;}
\tablefoottext{\rm *}{Sources with gas emission detected with ALMA;}
\tablefoottext{\rm **}{Derived for a dust temperature of 15~K.}
}  
\end{table*}

The ALMA observations were performed in 2012 on March 25$^{th}$, December 3$^{rd}$ and December 12$^{th}$ at Band~7, as part of the ALMA Cycle~1  program 2012.0.00542.S. 
From the 34 detections reported in BGH16, we obtained individual pointings with a total field of view (FoV) of $\sim\!\!18''$ for 30 of them (due to technical reasons, we did not observe the four sources LB01, LB32, LB33 and LB34). The central coordinates of each pointing correspond to the position of the pixel with maximum intensity of  the 30 APEX/LABOCA  detections. All the information about the 30 ALMA pointings  is included in Table~A1 of the Appendix.

A total of three data sets were collected, one for the northernmost region and two for the southernmost. We used 34 antennas (with $12\,{\rm m}$ diameter) in the northern region and 32 and 33 antennas in the South,  accounting for 3.7 hours of total integration time including overheads and calibration. Weather conditions were very good and stable,  with an average precipitable water vapour of 0.9mm (northern region) and 0.7-1.3 mm (southern region). The system temperature varied from 150 to $250\,{\rm K}$. 

The correlator was set up to four spectral windows in dual polarization mode, centered at 345.8 GHz, 347.8 GHz, 333.9 GHz, and 335.8 GHz (average frequency of 340.8\,GHz, or 880$\mu$m). The effective bandwidth used per spectral window was $1875{\rm MHz}$, providing a velocity resolution of  $\sim$ 0.42 km s$^{-1}$ after Hanning smoothing { providing an average $rms$ of 15\,mJy/beam}.

The ALMA calibration includes simultaneous observations of the $183\,{\rm GHz}$ water line with water vapour radiometers that measure the water column in the antenna beam, that is used to reduce the atmospheric phase noise.   Amplitude calibration was carried out using Ganymede and J0510+180. The quasars J$0532$$+0732$ and J$0607$$-0834$ were used to calibrate the bandpass, and J0521+1638 and J0539+1433 for the complex gain fluctuations.

Data reduction was performed using version 4.2.2 of the Common Astronomy Software Applications package (CASA). Imaging of the calibrated visibilities was done using the task CLEAN. We applied self-calibration using the continuum data in sources B30S-13 (LB19) and B30N-18 (LB14). Briggs weighting with robust parameter equal~2 was used in all continuum and line images presented in this paper.  The final images were corrected for the response of the primary beam and used to derive the physical parameters.

The average synthesized beam sizes of the continuum data in the North and South regions are 0\farcs93 $\times$ 0\farcs50 (PA=111 degrees) and 0\farcs95 $\times$ 0\farcs48 (PA=-63.5 degrees), respectively.  The shortest baseline of the ALMA observations is $\sim$ 15\,$k\lambda$, while the upper end of the baseline  range corresponds to $\sim$395\,$k\lambda$. Following equation A5 from \citet{Palau2010}, we estimate that the largest angular scale (LAS) detectable by the ALMA observations is $\sim$6".

We report the detection of five sources (out of 30 pointings) in our ALMA dataset. The five detected objects are spatially unresolved. Three sources are located in the northern region of B30, and two in the South, well within the ionization front facing the hot star $\lambda$~Ori  (see Figure~\ref{fig:wideALMA}). Their coordinates and fluxes are included in Table~\ref{table:almadets}, while the ALMA continuum images are displayed in Figure~\ref{fig:ALMAdet}. 
The separation of the ALMA detections from the phase centers are included in Table~1. As can be seen, LB31 is very close to the phase center and shows an  $rms$ of 0.13\,mJy. The other four sources are{ located at separations between 4.2 and 8.6 arcsec from the phase center} and display a larger $rms$  after correction from the response of the ALMA primary beam, which is $\sim$18 arcsec (see Table~\ref{table:almadets}).  We note that these separations are well within the APEX/LABOCA beam  (FWHM of 27.6") for all the ALMA detections.
In the five cases, there are no other detected sources in the ALMA FoV. In the remaining  25 pointings no sources were detected within the whole ALMA FoV.

Although the ALMA observations were designed to detect continuum emission, the spectral setup was also selected to detect possible gas emission in the CO(3-2) transition at 345.8\,GHz. The analysis of the spectral window centered at that frequency shows the presence of very extended gas emission, mainly resolved out, in the primary beam of most of the positions observed in the southern region of B30, and very faint gas emission inside the primary beam of some of the positions surveyed in the northern part of the cloud. The $v_{\rm lsr}$ of the CO(3-2) cloud emission is $\sim$11 km/s, consistent with one of the velocity components
derived by \citet[][9.43\,km/s]{Lang2000} in the average spectrum of the region.  The analysis of the ALMA data shows that we have detected gas emission in only two sources that are also detected in the continuum (see Table~\ref{table:almadets}). They are discussed in detail in Section~\ref{sec:analysis}.

Finally, we have estimated the relative positional accuracy of the observations using Eq.~1 from \citet{Reid1988}.
Considering the beam major axis ($\sim$0.9$''$), we derive relative positional accuracies between 0.2$''$ for SNR of 4.6 (LB31) and $\sim$5\,mas for a SNR  of 176 (LB14). 
The absolute position accuracy is smaller than the synthesized beam width.

\subsection{SABOCA observations}

We performed observations in continuum at 350\,$\mu$m  with the Submillimetre APEX Bolometer CAmera (SABOCA) installed at the Atacama Pathfinder EXperiment (APEX\footnote{This work is partially based on observations with the APEX telescope. APEX is a collaboration between the Max-Plank-Institute fur Radioastronomie, the European Southern Observatory, and the Onsala Space Observatory}) telescope. SABOCA is a 39-channel bolometer array with a 1.5$'$ field-of-view and a 7.8$"$ full-width at half-maximum (FWHM) beam per bolometer channel. The data were obtained under program C-087.F-0011A-2011, between July 27$^{th}$ and 29$^{th}$,  2011 . Weather conditions were excellent 
with pwv $\sim$0.4-0.7mm (average of 0.45 mm). Opacity at the zenith was calculated using skydips, yielding zenith opacity values between 0.9 to 1.4. Pointing measurements were taken regularly. The pointing uncertainly is estimated to be approximately 2$"$. 
The focus was verified at the beginning of each observation. 
Flux calibration was performed using Mars and Uranus as primary calibrators, and HL Tau, V883~Ori, VY~CMa and CRL618 as secondary calibrators. The absolute flux calibration uncertainty is estimated to be $\sim$30$\%$.  

Our map was centered at R.A.= 05:31:20.5, Dec=+12:05:45, comprising a region of $\sim8'\times4'$.  For processing the data we used the Bolometer Array Analysis Software (BOA\footnote{http://www.apex-telescope.org/bolometer/laboca/boa/}) and CRUSH\footnote{http://www.submm.caltech.edu/ sharc/crush/} 
\citep[see][]{Kovacs2008} software packages. Data reduction processing included flat-fielding, calibration, opacity correction, correlated noise removal and de-spiking. Every scan was visually
inspected in order to identify corrupted data. We used and optimized data processing to detect faint point-like sources (options -faint -deep in CRUSH). The final image was smoothed to a resolution of 10.6 arcsec. The total  on-source observing time was 4.5 hours and the $rms$ reached was $\sim$50 mJy at the central areas of the map.

We detected a total of 17 sources in the SABOCA map (see Figure~\ref{fig:wideALMA}) with SNR between 3.0 and 4.6.{ Their  coordinates correspond to the pixel of maximum emission (peak intensity)}, and are provided
in Table~\ref{table:sabocadets} together with the measured flux densities. All detections are point-like sources.

\begin{table*}\caption{SABOCA 350\,$\mu$m detections in the southern region of B30}\label{table:sabocadets}
{\tiny

\begin{tabular}{lllllllll}
\hline
 SABOCA &    RA       &   DEC  &     LABOCA                   & Peak       & rms    & SNR   & $A_{\rm v}$$^{\rm b}$ & IR  \\
source     &                &            &   source $^{\rm a}$    & intensity &         &                                         &      & counterparts?  \\
                 &  [h m s]  &  [$^\circ$ \,\arcmin \, \arcsec] & &        [Jy] &           [Jy/beam] &     & [mag] &  \\ 
                            \hline\hline
B30-SB01 & 05:31:37.68  & 12:06:13.7   &                         & 0.16     & 0.05  &   3.2  &   2.3  &       NO      \\
B30-SB02 & 05:31:36.09  & 12:05:24.2   &                         & 0.14     & 0.04  &   3.5  &   3.6 &      YES, 3, ??  \\
B30-SB03 & 05:31:32.41  & 12:04:16.7   &   B30-LB22     & 0.16     & 0.05  &   3.2  &   2.8 &     YES, LB22f, NOT MEMBER  \\
B30-SB04 & 05:31:31.75  & 12:04:18.6   &   B30-LB22     & 0.14     & 0.04  &   3.5  &   2.8 &     YES, LB22a, MEMBER   \\
B30-SB05 & 05:31:29.47  & 12:04:34.7   &   B30-LB23     & 0.15     & 0.05  &   3.0  &   3.0 &     YES, LB23a, MEMBER   \\
B30-SB06 &05:31:28.82  & 12:04:44.8    &   B30-LB23     & 0.14     & 0.04  &   3.5  &   3.1 &      NO    \\
B30-SB07 &05:31:29.57  & 12:05:37.7    &                         & 0.17     & 0.05  &   3.4  &   3.2 &     YES, , MEMBER   \\
B30-SB08 &05:31:27.94  & 12:05:31.3    &   B30-LB19     & 0.23     & 0.05  &   4.6   &   2.7 & YES, LB19a, MEMBER \\  
B30-SB09 &05:31:22.29  & 12:05:30.7    &   B30-LB20     & 0.16     &  0.05 &   3.2  &   1.7 & NO  \\
B30-SB10 &05:31:19.12  & 12:05:15.0    &                         & 0.15     &  0.05 &   3.0  &   2.4 & NO  \\
B30-SB11 &05:31:15.20  & 12:04:49.3    &                         & 0.16     &  0.05 &   3.2  &  3.4 & YES, 2, ?? \\
B30-SB12 &05:31:13.98  & 12:05:25.3    &   B30-LB27     & 0.19     & 0.05  &   3.8   & 4.1 & NO   \\ 
B30-SB13 &05:31:10.55  & 12:05:43.3    &                         & 0.15     & 0.05   &  3.0  &  4.2 &  YES, 1, ?? \\
B30-SB14 &05:31:11.39  & 12:06:24.2    &                         & 0.16     & 0.05   &  3.2  &  3.9 &  YES, 1, ?? \\
B30-SB15 &05:31:07.69   & 12:06:30.9   &                         & 0.16     & 0.05   &  3.2  &  2.6 &  YES, 1, ??  \\
B30-SB16 &05:31:08.45   & 12:05:07.3   &   B30-LB28     & 0.22     & 0.05   &  4.4  &  3.8 & YES, LB28a, MEMBER \\
B30-SB17 &05:31:04.98   & 12:05:25.3   &                         & 0.22     & 0.07   &  3.1  &  1.8  & NO  \\ 
\hline
\end{tabular}
}
\tablefoot{ 
\tablefoottext{\rm a}{LABOCA designation number (from BGH16); }
\tablefoottext{\rm b}{From the 2MASS extinction map; }
\tablefoottext{\rm c}{In this column we specify if the SABOCA source shows counterparts in the IR, the number of counterparts (or its name if included
in BGH16), and their nature: member/not member of B30 (according to BGH16), or unknown (??).}
}
\end{table*}

\subsection{Complementary data}

In addition to the ALMA and SABOCA observations presented here, we used ancillary data of B30 to complement our observations.
The data are fully described in  BGH16. Briefly, we gathered deep, multi-wavelength observations of B30 
from the optical to the far-IR regime. The optical data were obtained with the Wide Field Camera (WFC) at the Isaac Newton Telescope in La Palma in two filters, $r$ and $i$ (INT/WFC), while deep near-IR data 
were collected with  Omega~2000 at the Calar Alto Observatory (CAHA/O2000).  The mid-IR data were obtained with the Spitzer space telescope with  both  the IRAC and MIPS instruments, covering a wavelength range from 3.6$\mu$m to 70$\mu$m.
We also complemented our database with archival observations from the infrared observatory WISE, far-IR data from the AKARI mission, and the {\em Planck} Catalog of compact sources (Release~1) . Finally, we 
also considered the 870\,$\mu$m LABOCA data included in BGH16. The pointing errors, completeness and 
limiting magnitudes of all these data (except $Planck$) are fully described in BGH16.

\begin{figure}[t!]
\includegraphics[width=0.5\textwidth,scale=0.5]{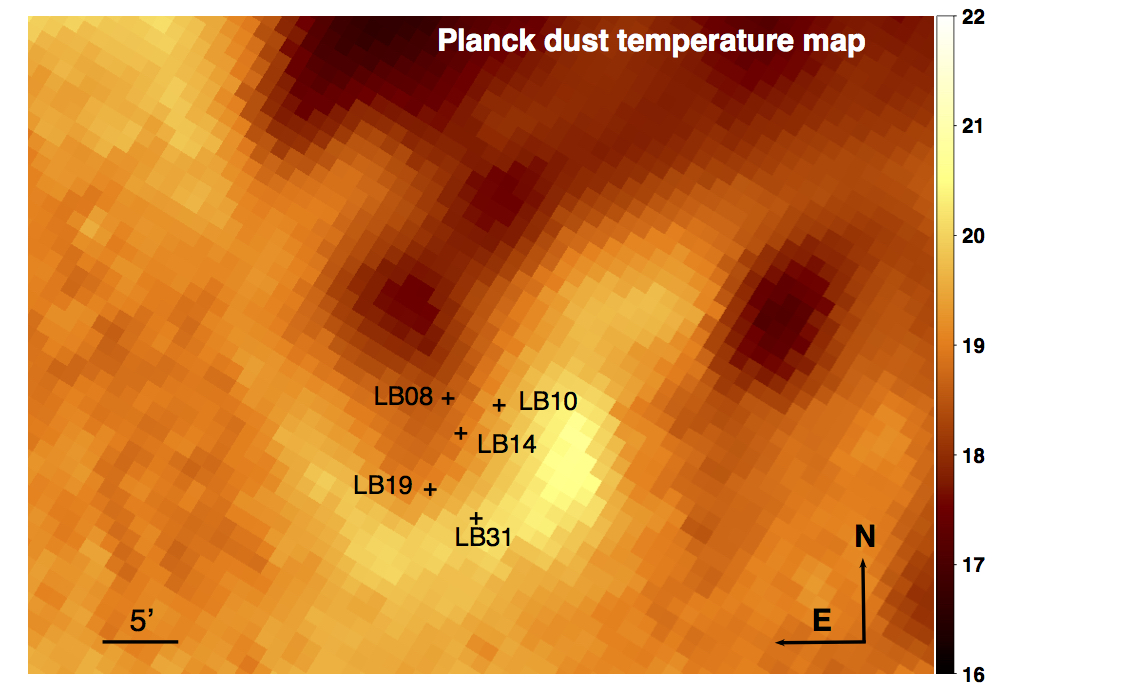}
\caption{Dust temperature map of the B30 region (FoV of $\sim$1$^{\circ}$$\times$ 0.7$^{\circ}$) from the {\em Planck} mission \citep{Planck2014}. The color bar displays the temperature values in K. We have included the positions of the detected ALMA sources.}\label{planck}
\end{figure}

\begin{figure*}[t!]
\center
\includegraphics[width=0.9\textwidth,scale=0.5]{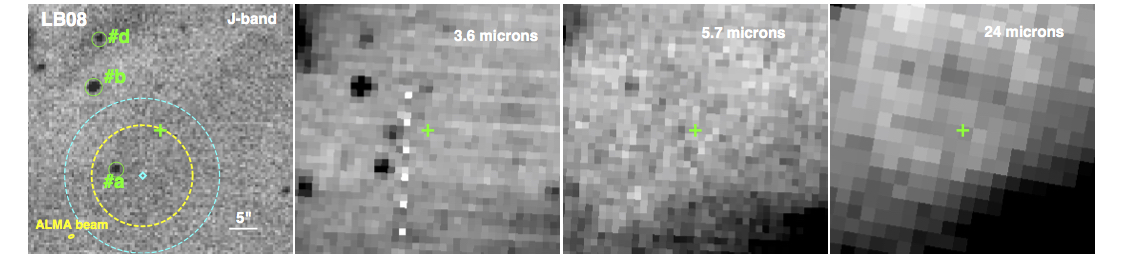}
\includegraphics[width=0.9\textwidth,scale=0.5]{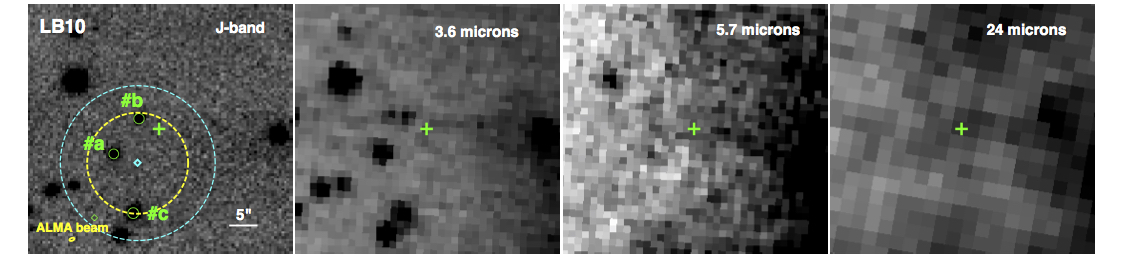}
\includegraphics[width=0.9\textwidth,scale=0.5]{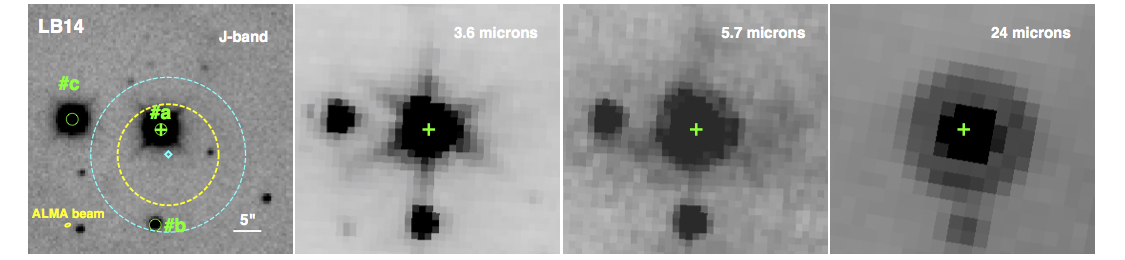}
\includegraphics[width=0.9\textwidth,scale=0.5]{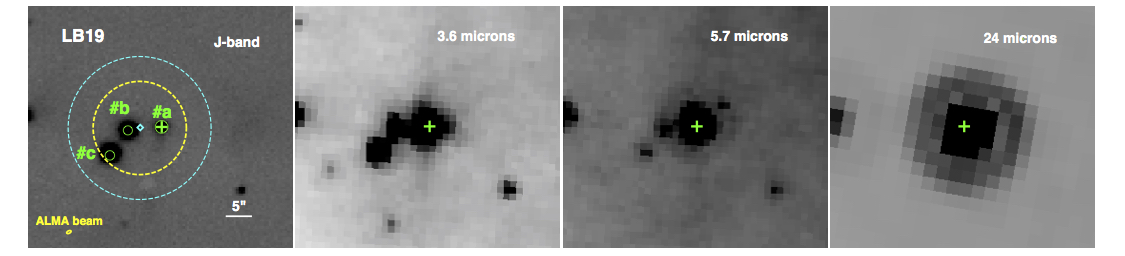}
\includegraphics[width=0.9\textwidth,scale=0.5]{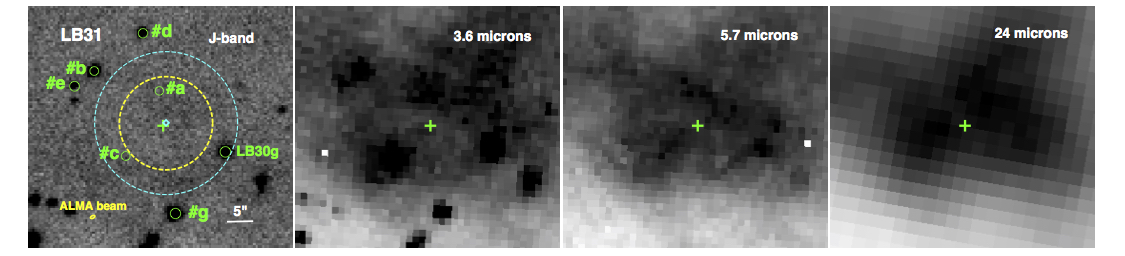}
\caption{\label{alma_fcs} 
Infrared (CAHA/O2000 and Spitzer/IRAC and MIPS) images centered on the ALMA detection coordinates. North is up and East to the left. We have included in all the J-band images (left panels) the position of the ALMA detections (green crosses), and the possible infrared counterparts to the APEX/LABOCA detections identified by  BGH16 (green labels). The  cyan diamonds represent the APEX/LABOCA peak intensity coordinates (ALMA phase center coordinates).  { We have also plotted the APEX/LABOCA beam of 13.8" radius (cyan dashed circle), the ALMA FOV of $\sim$18\arcsec diameter (yellow dashed circle)}, and the  ALMA beam at the bottom left corner (yellow solid ellipse).}
\end{figure*}

We also used an extinction map of the B30 cloud based on 2MASS J-band data and derived using the 
star-count method presented in \citet{Cambresy1997}. We estimated $A_{\rm v}$ from $A_{\rm J}$ using the relations by \citet{Fitzpatrick1999}. 
The values of $A_{\rm v}$ at the position of the ALMA and SABOCA detections 
are provided in Tables~\ref{table:almadets} and \ref{table:sabocadets}, respectively.
We note that the average spatial resolution of the extinction map is 1\farcm5, and the typical uncertainties are of $\sim$0.1\,mag.

\section{Data Analysis \label{sec:analysis}}

\subsection{ALMA detections: continuum and gas emission}

As mentioned above, we detected five sources with ALMA, all of them spatially unresolved.
Two of these sources, LB14 and LB19, also show gas emission.

If we assume that the dust emission is optically thin, we  can estimate the total mass (gas and dust) from 
thermal continuum emission following the expression:

\begin{equation}
M = \frac{S_\nu\; D^2}{B_\nu (T_d)\; \kappa_\nu}
,\end{equation} 

\noindent
where $S_{\nu}$ is the flux density, $D$ is the distance to the source,  $B_{\nu}$($T_d$) is the Planck function 
at the dust temperature $T_d$, and $\kappa_\nu$ is the absorption
coefficient per unit of total mass density.
We can rewrite Eq.~1 in this way:

\begin{equation}
\Bigg[\frac{M}{M_\odot}\Bigg]= 3.25 \;\frac{e^{0.048\nu/T_d}-1}{\nu^3 \kappa_\nu}\;\Bigg[\frac{S_\nu}{\rm Jy}\Bigg]\Bigg[\frac{D^2}{\rm pc^2}\Bigg]  
.\end{equation}

\noindent

The dust temperature maps generated by the $Planck$ mission \citep{Planck2014} show temperature values between 17 and 21\,K in the B30 region (see Figure~\ref{planck}). This temperature can be considered as an upper limit to the dust temperature of the starless cores of B30, where the temperature should decrease towards the core centers \citep[for example, the starless core in the Pipe Nebula FeSt\,1-457 presents a decrease in dust temperature from $\sim19$~K in the core outskirts down to $\sim14$~K in the core center,][]{Forbrich2015}. On the other hand, the Planck temperature should be considered as a lower limit to the dust temperature for cores already harbouring protostars. Therefore, we have selected a conservative dust temperature of 15\,K to estimate the masses of the cores in B30. We have estimated the absorption coefficient for an average frequency of 340.8 GHz ($\sim$880\,$\mu$m) 
interpolating  from the tables of \citet{Ossenkopf1994} for the case of thin ice mantles and a density of 10$^6$\,cm$^{-3}$. We obtained a $k_{\nu}$ of 
0.01715 cm$^2$g$^{-1}$ for a gas to dust mass ratio of 100. Finally, we have assumed a distance of 400\,pc, and the ALMA flux densities included in Table~\ref{table:almadets}. The estimated masses are included in Column~7 of Table~\ref{table:almadets}. They range from  $\sim$  0.9 to 66\,$M_{\rm Jup}$.
We note that the uncertainty in the masses due to the opacity law and dust temperature is estimated to be a factor of four.

The comparison of  the masses derived from ALMA data with those derived from APEX/LABOCA data shows 
much smaller values for the ALMA detections (see Columns~7 and~8 in Table~\ref{table:almadets}).
We have therefore compared the fluxes from  the two datasets and derived the percentage of  flux that ALMA did not recover
compared to a single-dish (LABOCA). This is provided in Column~9 of Table~\ref{table:almadets}.
We conclude that for all the sources except LB14, ALMA is missing $>$ 85\% of the flux detected by LABOCA.

\begin{figure*}
\includegraphics[width=1.0\textwidth,scale=0.40]{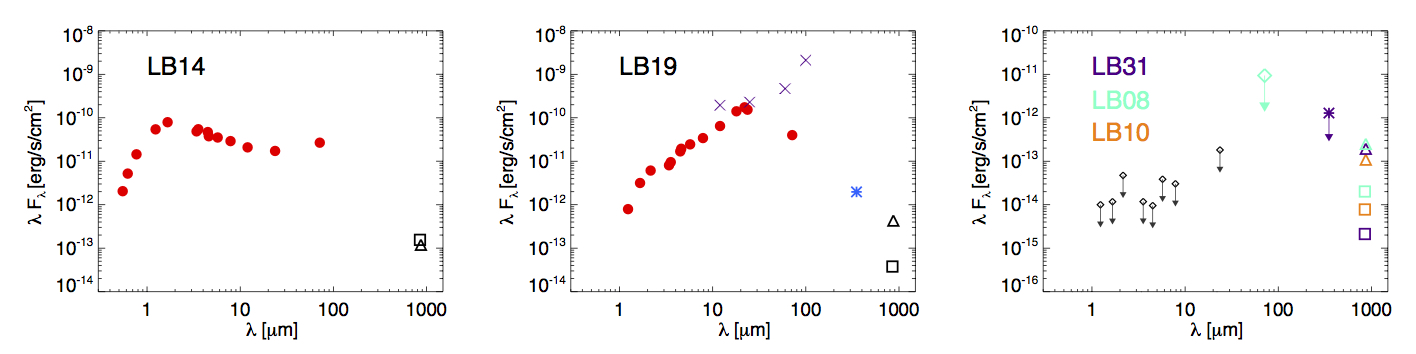} 
\caption{\label{alma_sed} Observed SEDs of the five ALMA detections. In all the panels, the open triangles and squares represent the APEX/LABOCA and ALMA fluxes, respectively, while the asterisks represent the APEX/SABOCA data. {\em Left and middle panels:} We show the complete SEDs of LB14 and LB19.
The red circles represent optical to mid-IR data from all our catalogues whenever available (see section 2.3). 
 The purple crosses in the LB19 panel represent IRAS data.  
{\em Right panel:} Three ALMA detections with no infrared counterparts: LB08, LB10 and LB31. The APEX/LABOCA and ALMA 
detections are color-coded for each source.  We have  included the limiting magnitudes (black diamonds) of our infrared data. }
\end{figure*}

We looked for possible infrared/optical counterparts to the ALMA sources using a search radius of 1" (the ALMA beam) and 
found clear infrared counterparts for two out of five ALMA detections (see Figure~\ref{alma_fcs}): LB14 and LB19. 
The counterparts are the sources named LB14a and LB19a in BGH16, and were  detected 
within the inner 5" of the LABOCA beam. Their SEDs are included in Fig~\ref{alma_sed}.

Three ALMA sources, namely LB08, LB10 and LB31 did not show any counterpart within 1" radius. BGH16 reported 
infrared counterparts (namely LB08a and LB10a) within a 5" radius for the two LABOCA cores  (LB08 and LB10)
but the positional accuracy of ALMA (and the CAHA and Spitzer telescopes) is good enough to reject them as the 
880$\mu$m sub-mm emitters. For these three objects  we also searched for counterparts in the NASA/IPAC 
Extragalactic Database (NED), but did not find any object within a 5" radius of the ALMA position.

The individual ALMA detections are discussed in more detail in the following subsections. We begin by describing the sources with IR counterparts
and end by describing those without them.

\begin{figure*}
\includegraphics[width=1.0\textwidth,scale=0.5]{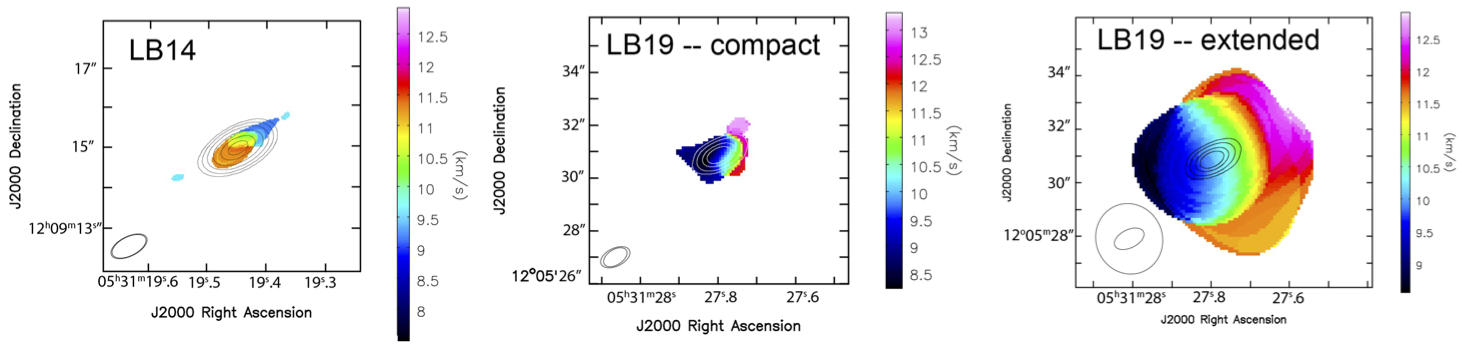}
\includegraphics[width=1.0\textwidth,scale=0.5]{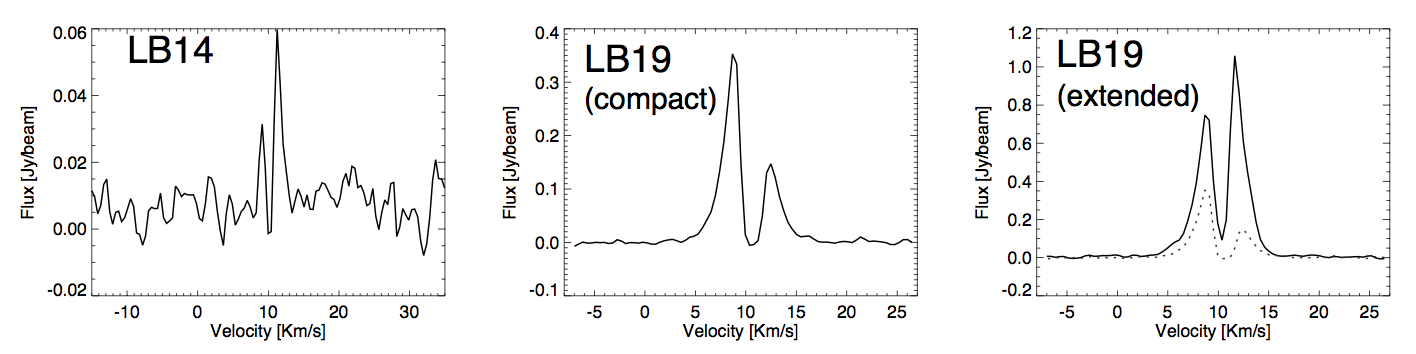}
\caption{\label{alma_gas} 
{\bf Top panels:} 
ALMA CO(3-2) intensity weighted velocity field (first order moment, color scale) with the ALMA 880~micron continuum emission over-plotted in contours. 
The left panel corresponds to LB14, while the middle and right panels correspond to the compact and extended emission of LB19, respectively. 
For LB14, contours are 5, 10, 20, 40, 80, and 120 times the rms noise of the map, 0.25 mJy/beam.
For LB19, contours are 5, 10, 20, and 30 times the rms noise of the map, 0.20 mJy/beam. 
Ellipses at the bottom left represent the synthesized beam for the continuum and for the CO(3-2) maps. In the case of LB19, the beam for the CO(3-2)  image corresponds to emission at $>$ 60 $K\lambda$ (central panel), and $<$ 60$K\lambda$ (right panel).
{\bf Bottom panels:} CO(3-2) spectrum in the region of emission of LB14 (left),  and in the compact (middle) and extended (right) 
regions of emission of LB19. In the last panel both the compact (dotted line) and extended emission (solid line) are superimposed for comparison.
}
\end{figure*}

\subsubsection{LB14\label{sub:lb14}}

LB14 has a very bright counterpart at optical and infrared wavelengths, named \#a in BGH16, and classified as a
Class~II stellar member (see Figure~\ref{alma_fcs}).  The ALMA detection is clearly assigned to this infrared source, with a separation of 0\farcs3. We note that this object was already classified by \citet{Dolan2002.1} as a pre-Main Sequence (PMS) star.

We have compared the optical photometry of the target \citep[$R$=16.016$\pm$0.032, $I$=14.486$\pm$0.032\,mag,][]{Dolan2002.1}, corrected by the visual 
extinction provided in Table~\ref{table:almadets}, with BT-SETTL evolutionary models \citep{Allard2012.1}. We obtained that
the central star shows a mass  of $\sim$0.3\,M$_{\odot}$ (or $\sim$0.5\,M$_{\odot}$) for an age of 1\,Myr (or 3\,Myr).

The IRAC colors of LB14a, [3.6]-[4.5]=0.56 and [5.8]-[8.0]=0.80 mag, are consistent with  a Class~II object \citep{Allen2004}. 
The SED is  displayed in Figure~\ref{alma_sed} (left panel)  showing a clear excess at long wavelengths.
The object is not included in the field of APEX/SABOCA.

The source is point-like in both the APEX/LABOCA and ALMA data. In fact, 
the estimated sub-mm fluxes are very similar in the two datasets
(35$\pm$14\,mJy versus 44.0$\pm$0.5 mJy, assuming 2-$\sigma$ errors), with estimated masses 
of $\sim$ 51\,M$_{\rm Jup}$ and $\sim$67\,M$_{\rm Jup}$, respectively, and 0\% of ALMA missing flux when compared to the 
APEX/LABOCA emission.
Given the evolutionary stage of LB14, we expect that the estimated mass is mainly associated with a circumstellar disk. In fact, since this object is evolved in comparison with starless cores, the dust temperature should be higher than the assumed 15\,K. We estimate a disk mass of 43\,M$_{\rm Jup}$ for 20\,K, or 24\,M$_{\rm Jup}$ for 30\,K.  The angular resolution of the ALMA data provides an upper limit to the disk radius of $<$140\,AU. 

LB14 shows gas emission in the ALMA data. 
The CO(3-2) emission is very compact, spatially unresolved, and centered at the position of the dust emission (see Fig~\ref{alma_gas}, left top panel). The gas emission is seen from 8.3 to 12.1 km/s.  The spectrum shows a double peak emission (Fig~\ref{alma_gas}, left bottom panel) suggesting a disk in rotation, with the red-shifted emission located to the south-east and the blue-shifted one to the north-west of the beam center. The spectrum is centered at a $V_{\rm lsr}$ of 10.4 km/s, confirming this source as a member of B30.

\subsubsection{LB19\label{subsec:lb19}}

In the case of LB19, the ALMA detection is associated with source \#a in BGH16 (see Figure~\ref{alma_fcs}).
The source is undetected in the optical images, but detected in the J-band with a
brightness that increases towards longer wavelengths. The source has a clear spatially resolved counterpart 
in the Spitzer images. Its IRAC colors, [3.6]-[4.5]= 1.34, [5.8]-[8.0]=1.37 are consistent with a Class~I source \citep{Allen2004}.

In  the APEX/LABOCA map, LB19 is detected as an extended source with a 
deconvolved angular size of 24". The target is also detected in the APEX/SABOCA map and looks slightly elongated. However,
the low SNR  of the 350\,$\mu$m data does not allow its confirmation as an extended source.

The SED of the object is displayed in the central panel of Figure~\ref{alma_sed} and its shape resembles that of a Class~I object.  In fact, LB19 is associated with 
the IRAS source IRAS05286$+$1203 (RA=82.8671, DEC=+12.0899, separation of 0.2\arcsec) included in the near-IR survey of Class~I protostars 
by \citet{Connelley2008}. They estimated a $L_{\rm bol}$  of 14.4\,L$_{\odot}$ considering only the emission from the four IRAS bands. 
The source has been detected with Spitzer/MIPS  at 70$\mu$m and displays a significantly smaller flux  than in IRAS 60 $\mu$m (0.94\,Jy vs. 9.35\,Jy), with the
latter value displaying an IRAS quality flag of '2' (moderate quality). 
If we integrate the whole SED without considering the IRAS data-points,  we obtain a $L_{\rm bol}$ of 1.2\,$L_{\odot}$ and a temperature of $T_{\rm bol}$ $\sim$ 235\,K, which is consistent with a low-mass protostar. 

The estimated mass from the ALMA data is $\sim$16\,M$_{\rm Jup}$, while the estimation from APEX/LABOCA data is 
182\,M$_{\rm Jup}$ considering all the integrated emission, or 116\,M$_{\rm Jup}$ if we only consider the peak intensity.
Such a difference is probably related to the APEX/LABOCA measurement (peak intensity of $\sim$79 mJy) 
including a significant emission from the surrounding cloud.

LB19 is the second source with a gas detection.
The CO(3-2) emission shows two components: one that is compact and spatially coincident with the continuum detection, and an extended component (baselines < 60 k$\lambda$) of  $\sim$2400 AU surrounding it (Figure~\ref{alma_gas}, top middle and right panels). Both compact and extended components show emission from 5.7 to 14.2 km/s and are centered at 10.4 km/s, which confirms LB19 as a B30 member. 
The intensity-weighted  velocity field of both compact and extended components (first-order moment) shows a velocity gradient with the most blueshifted and redshifted material at the tips of the gradient.

The spectrum of the compact component is asymmetric,  with the blueshifted side stronger than the redshifted one (see Fig~\ref{alma_gas}, top middle panel). This might suggest infall is occurring at small scales ($\sim$400 AU) in this source. 

The extended component is slightly elongated in the south-east north-west direction and shows also extended emission close to the cloud velocity 
($\sim$11.5~km/s) in the south-west north-east direction.  There is no presence of high velocity wings related to molecular outflows. The kinematical pattern might suggest that we are observing a rotating oblate envelope with probable infalling at 400\,AU scales near the central protostar.  The extended component does not show a clear infalling pattern.

\begin{figure}[h!]
\includegraphics[width=0.5\textwidth,scale=0.5]{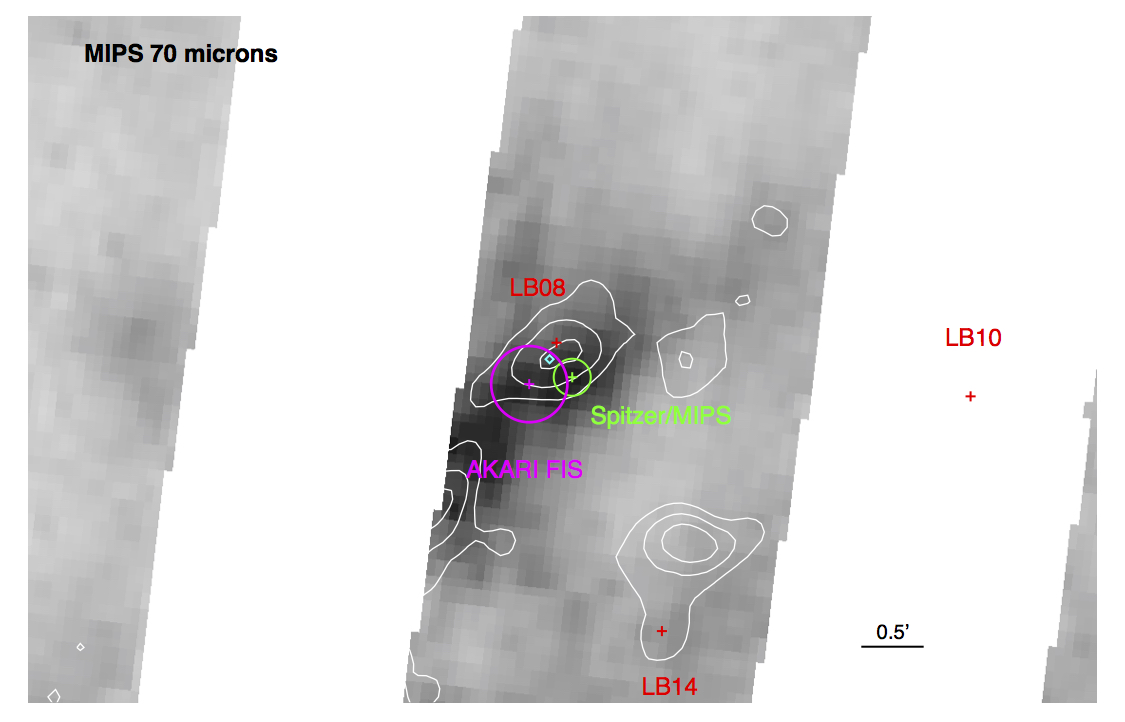}
\caption{Spitzer/MIPS 70\,$\mu$m sub-image centered on the ALMA source LB08 within B30. The white regions do not have data available. We overplotted the LABOCA contours (white), the ALMA detections (red), and the AKARI/FIS detection  (magenta).
The magenta circle represents the FWHM PSF of the AKARI FIS data at 65\,$\mu$m (37'' diameter). The source LB08 (red cross) is within a bright filamentary structure detected at 70 microns and extending from the Northwest to the Southeast. The MIPS/70\,$\mu$m peak emission (green cross) is located at $\sim$18 arcsec SW from the ALMA detection, and the green circle represents the FWHM of the PSF  (18" diameter for MIPS).}\label{fig:mips70}
\end{figure}

\subsubsection{LB08, LB10, and LB31}\label{preBD}
                             
LB08, 10, and 31 are the weakest sources detected with ALMA. There is no presence of gas emission at their 
positions, with an $rms$ per channel of 13 mJy/beam, 11 mJy/beam, and 18 mJy/beam, respectively.

While LB08 and LB10 lie in the  northern  region of B30,  LB31 is located in the southern part of the cloud,  well within the
ionization front seen in the 24\,$\mu$m image (see Figure~\ref{fig:wideALMA}). While LB08 shows extended emission in the APEX/LABOCA map, 
LB08 and LB10 are detected as point-like sources. None of these ALMA detections show a 
clear counterpart in the infrared regime (Figure~\ref{alma_fcs}).  

LB08 shows  a deconvolved angular size of 27" in the APEX/LABOCA data after fitting a Gaussian.
BGH16 identified an infrared counterpart within the inner 5" of the LABOCA beam: LB08a (see Figure~\ref{alma_fcs}).
However,  the ALMA absolute positional accuracy is good enough to conclude that  this  infrared source is unrelated with the ALMA detection. We note that there is  a detection with AKARI/FIS at 65, 90, 140 and 160$\mu$m at a separation of  
24" from the ALMA source (AKARI source name 0531239+121115). The AKARI/FIS PSF FWHM at the four different bands  are 37, 39, 58 and 61\,arcsec, respectively, with a positional accuracy of $\sim$8\,arcsec \citep{Yamamura2009}. 
The Spitzer/MIPS image at 70\,$\mu$m does not cover the whole field of B30, and displays stripes without data every 3' (see Figure~\ref{fig:mips70}). However,  the region of LB08 is included and its surroundings show bright extended emission close to the ALMA detection. The bulk of this emission  (the MIPS peak intensity is at 18" SW from the ALMA detection) is located more to  the South, tracing a filament along the northwest-southeast direction (the MIPS PSF FWHM at 70 microns is of 18"). In fact, the AKARI/FIS detection is at the center of this region and it 
probably includes most of the emission of the filamentary structure. We therefore conclude that AKARI detection
seems not to be associated with the ALMA source but, alternatively, with the filament. On the other hand, the fact that the ALMA source is  detected in this filament 
 reinforces the hypothesis that it belongs to the B30 region and is not a background contaminant.
 
In the case of LB10, BGH16 found an infrared counterpart (source \#a) within 5 arcsec of the LABOCA beam (see Figure~\ref{alma_fcs}).
The ALMA data show that the detection is clearly unrelated to \#a and, in fact, is closer to source \#b, 
classified as a B30 non-member by BGH16. The separation between the ALMA detection and source \#b is of $\sim$4", therefore they seem unrelated.
Inspection of the {\em Spitzer}/IRAC and MIPS images does not reveal any counterpart at the position of the ALMA source either.

Finally, LB31 is detected with ALMA with a SNR of $\sim$4.6.  BGH16 did not find any IR counterparts within 5 arcsec of the sub-mm LABOCA source, and we do not identify any counterpart at the position of the ALMA source. We only detect extended emission from the B30 cloud at  24$\mu$m, since the object is well within the ionization front of the cloud (see Figure~\ref{fig:wideALMA} and a zoom of the LB31 region in Figure~\ref{alma_fcs}). 
This is the only object (out of the three) that is included in the  SABOCA map, but it is not detected at 350\,$\mu$m.  

Note that neither LB10 nor LB31 have 70$\mu$m MIPS data available, and they are not detected with AKARI.
As mentioned above, neither of them appears to have a counterpart in the NED database.


\subsection{APEX/SABOCA detections}
\begin{figure*}
\includegraphics[width=1.0\textwidth,scale=0.5]{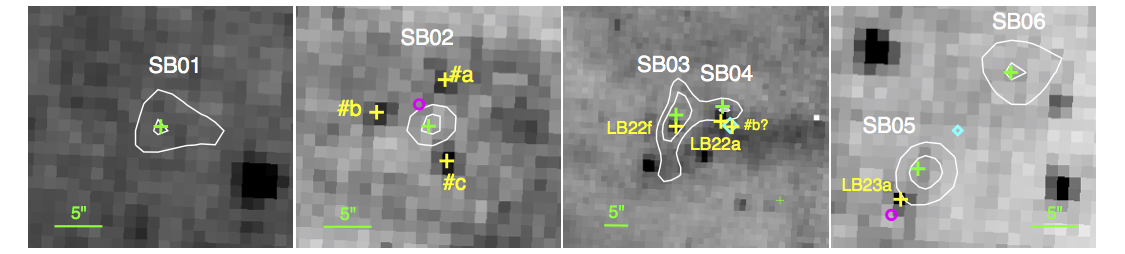}
\includegraphics[width=1.0\textwidth,scale=0.5]{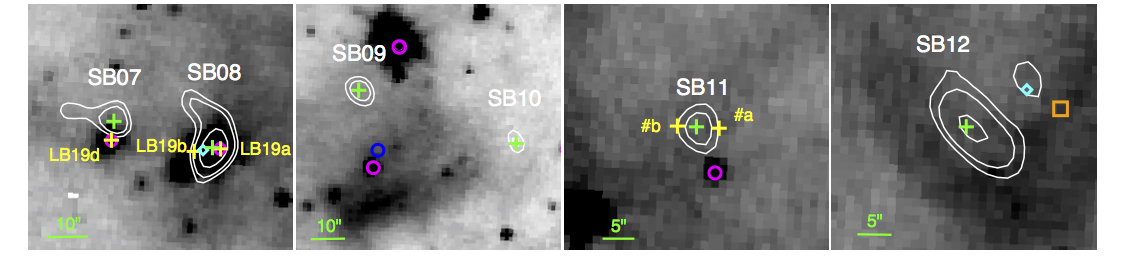}
\includegraphics[width=1.0\textwidth,scale=0.5]{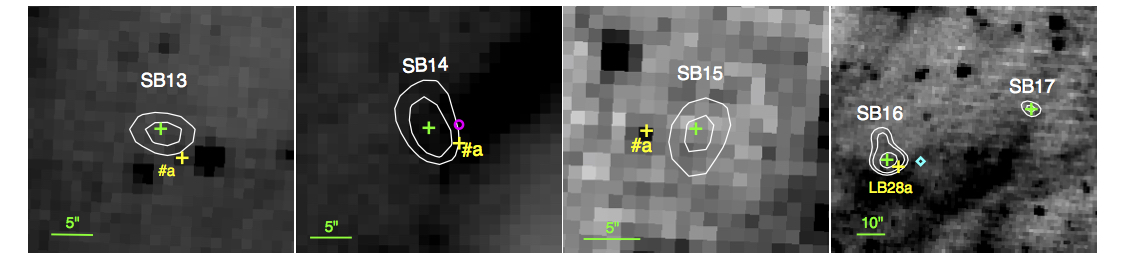}
\caption{\label{saboca_fcs} 
Spitzer IRAC 3.6\,$\mu$m finding charts with the APEX/SABOCA detections, except for SB14 (IRAC 8\,$\mu$m image).
North is up and east to the left. The green crosses represent the coordinates of the  peak emission of the APEX/SABOCA detections.
The white contours represent the APEX/SABOCA emission at 2.5, 3, 4 and 5-$\sigma$ level. The infrared counterparts within a radius 
of 5\farcs3 (half of the APEX/SABOCA beam) are represented  by yellow crosses. If previously discussed by BGH16, they are named 
with the APEX/LABOCA designation. If not, they are named with low-case letters. The magenta circles represent WISE detections, the cyan diamonds represent the
position of the APEX/LABOCA cores peak intensity, and the blue circles indicate $AKARI$ detections. The orange square represents a $Planck$ detection (see text).}
\end{figure*}

We detected a total of 17 sources in the APEX/SABOCA map at 350\,$\mu$m (see Figure~\ref{fig:wideALMA}). 
Eight out of the 17 seem to be associated with LABOCA cores:
sources B30-SB03 to 06 are contained within the  southern filament  revealed in the LABOCA contour map.
While SB03 and SB04 seem to be related to B30-LB22, B30-SB05 and B30-SB06 are associated with B30-LB23 (Figure~\ref{fig:wideALMA}).
B30-SB08 is related to B30-LB19 (already discussed in section~\ref{subsec:lb19}).
B30-SB09, at the center of the map,  is at $\sim$9 arcsec from B30-LB20.  In the western 
part of the map we find B30-SB12, at 10.5" from the bright core B30-LB27, and B30-SB16,  at 11.5" from  B30-LB28.
The low SNR of the detections has prevented us from estimating the dust masses.

\begin{figure}
\includegraphics[width=0.5\textwidth,scale=0.5]{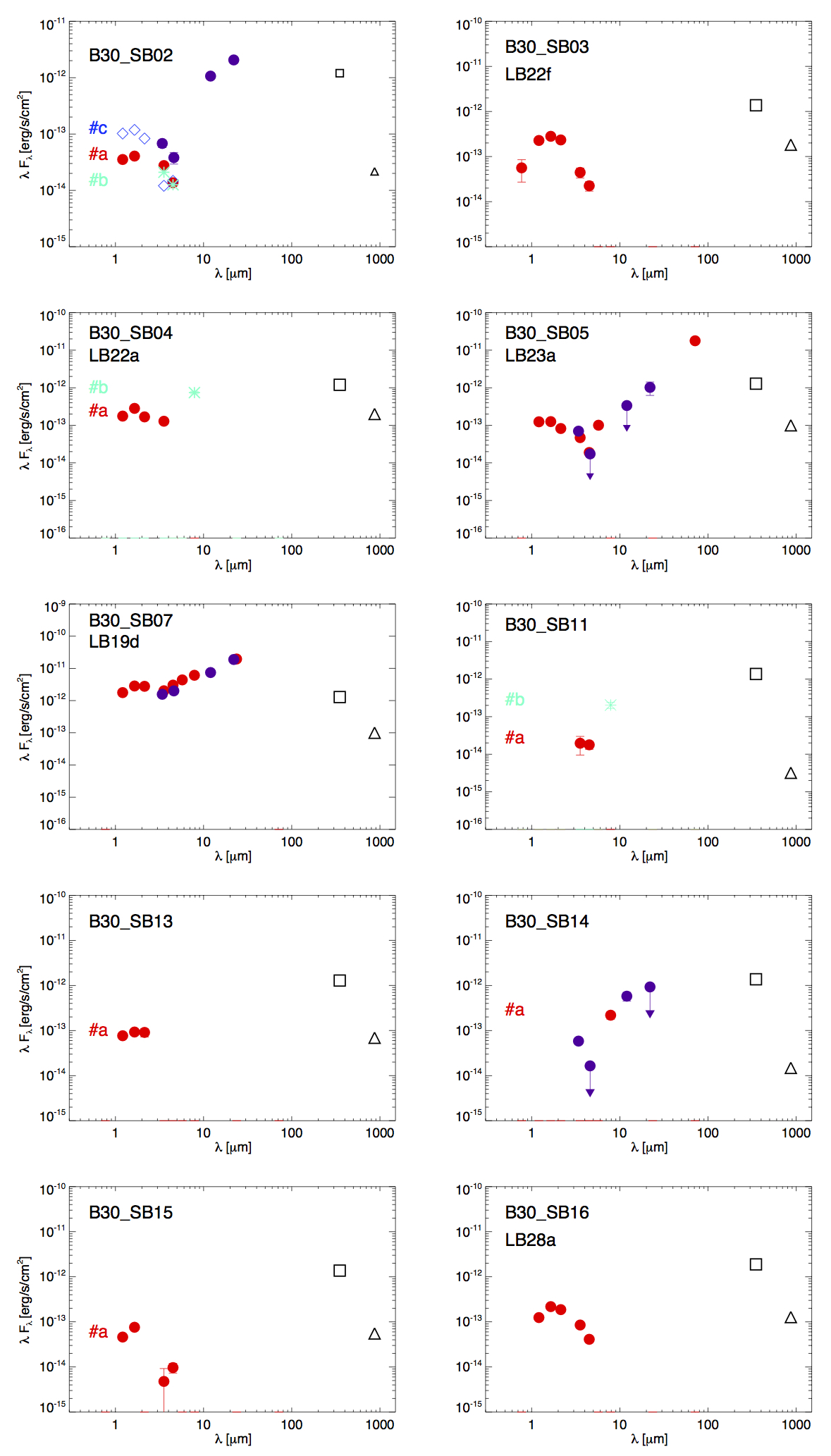}
\caption{\label{saboca_seds} SEDs of the infrared counterparts to the APEX/SABOCA detections. 
Those with multiple counterparts are plotted with different colors.  The purple circles represent WISE data 
available for some of the sources.
}
\end{figure}

We looked for infrared counterparts to all the SABOCA sources in our Spitzer/IRAC and CAHA/O2000 catalogues. We  used a diameter search of 10\farcs6 which corresponds to the SABOCA beam. We also looked for possible counterparts with WISE\footnote{http://wise2.ipac.caltech.edu/docs/release/allwise/}
and AKARI. The finding charts are displayed in Figure~\ref{saboca_fcs}, and the SEDs of the detected counterparts in Figure~\ref{saboca_seds}.
For the eight sources that seem associated with LABOCA cores, we have refined the assignment of infrared counterparts from  BGH16. 
For the  SABOCA sources not associated with LABOCA cores, we can see that   
they all lie  within the southern filament detected in the LABOCA map (excluding source 17, at the edge of the map), 
and are associated with some 870\,$\mu$m emission at a $\geq$ 3-$\sigma$  level (sources 02, 07, 10, 13, and 15, see Figure~\ref{fig:wideALMA}) 
or with a more marginal ($<$ 3.0-$\sigma$) emission (sources 01, 11, and 14).
For these sources we have represented  the LABOCA peak emission at the SABOCA position in their SEDs. 

We found infrared counterparts for 11 SABOCA sources. However, only one of them (LB19a) is detected close to the center of the SABOCA beam
at a separation of 1.7 arcsec. For the ten remaining sources, the assignment of an IR counterpart is uncertain, but we describe our findings below for
completeness (to facilitate the reading, we have omitted the prefix B30 in the APEX/SABOCA source names):

For  {\bf SB02} we detected three possible Spitzer/IRAC counterparts, all of them at separations  between 4" and 5". 
Sources \#a and \#c are also detected in the near-IR. 
SB02 has a single WISE counterpart at a separation of 1.3 arcsec (WISE J053136.15+120526.4).
Given its smaller spatial resolution, we think that the WISE source might contain emission from the three sources detected by Spitzer.
We cannot assign the correct counterpart with the available data, although the final SED resembles that of a young object.

\begin{figure}[t!]
\includegraphics[width=0.5\textwidth,scale=0.5]{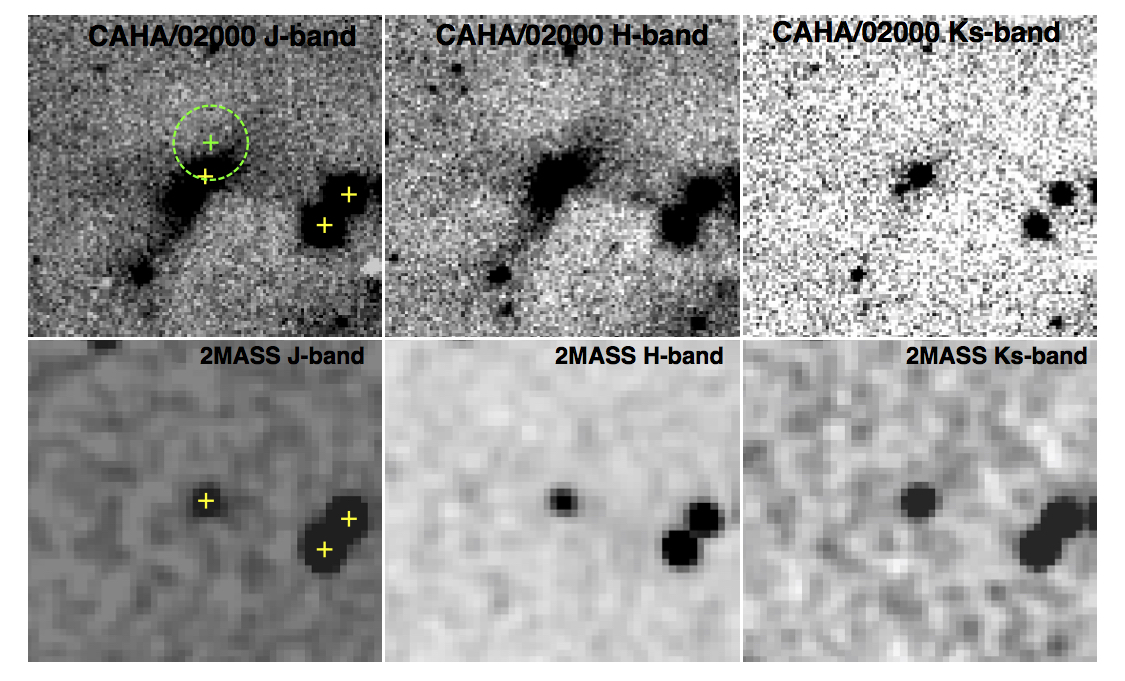}
\caption{Infrared images of the source SB07 (LB19d). North is up and East to the left. The upper panels show our CAHA/O2000 $JHK_s$ images,  and the lower ones the 2MASS images. We spatially resolve the object into two sources. The $J \& H$-band images show diffuse extended emission resembling a bipolar nebula.{ In the upper left panel we have
represented the  peak intensity coordinates of the SABOCA source SB07 (green cross), the SABOCA beam size (10.6" diameter) and the 2MASS detections in the field (yellow crosses in the J-band images). }}\label{LB19d}
\end{figure}

{\bf SB03 and SB04} appear to be associated with B30-LB22, although SB04 is closer ($\sim$4 arcsec) to the LABOCA peak than 
SB03 ($\sim$11"). In the case of SB03, the only possible detected counterpart is LB22f, while 
 we can assign LB22a to SB04 (following the nomenclature by BGH16).
LB22f is classified as a non-member by BGH16 and its SED  does not, in fact, display any infrared excess. 
Therefore, we can assume that the infrared source is unrelated to SB03, which could be classified as a starless core. 
LB22a is classified as a B30 member with a peculiar SED (with only four points, see  Figure~\ref{saboca_seds}). The target is detected
in CAHA/O2000 and  IRAC/3.6\,$\mu$m at a separation of 2 to 3 arcsec from the SABOCA source. 
Finally, we note that the 8$\mu$m IRAC image shows extended emission close to  SB04, with a tentative  infrared detection  at 4.7 arcsec (source \#b? in Figure~\ref{saboca_seds}).

 The two SABOCA sources {\bf SB05 and SB06 }are close (5.8" and 8.0", respectively) to the peak of the LABOCA core B30-LB23. SB05 is at a separation of 3.8 arcsec from the infrared source LB23a, 
which is classified as a probable member by BGH16. Its SED shows a clear IR excess starting at $\sim$6\,$\mu$m. We also detect a source  with WISE slightly further away (5.6" from the SABOCA peak coordinates, WISE J053129.66+120429.8) but,  given the WISE angular resolution (6.1" at 3.4\,$\mu$m), is probably related with LB23a. 
However, we think that if the infrared source LB23a (which is very compact) were related to the SABOCA and LABOCA emission, it should have been detected with ALMA when pointing at LB23, but it was not.  Therefore, we can conclude that they are probably unrelated.  Since the ALMA LAS is 6", the SABOCA and LABOCA emission have to be associated  with a structure larger than 6" (or 2400\,AU at 400\,pc). In fact, it is also possible that the Spitzer/MIPS 70 $\mu$m emission is also unrelated 
to the compact infrared source but associated to  SB05 and LB23.
The source SB06  does not show any infrared counterpart, and can  be considered a starless core.

For {\bf SB07} we detect an IR counterpart  at a separation of 4.9 arcsec. This is a very bright source close to B30-LB19, and named  LB19d by BGH16. It is classified as a 
member of B30, and its SED shows a clear infrared excess. Curiously, this object is included in the NED database and classified as a galaxy ('G'), based on 
its inclusion in the 2MASS Catalogue of extended sources. To the best of our knowledge there is no other evidence pointing towards an extragalactic nature according to the NED database. As seen in Figure~ \ref{LB19d}, the object  is resolved into two  sources in our CAHA/O2000 near-IR  infrared images, and displays some diffuse extended emission that resembles that of a bipolar nebula. The infrared counterpart LB19d is also detected  with  WISE (J053129.61+120532.8) and the final SED is typical of a young stellar object. 


 As already discussed in Section~\ref{subsec:lb19}. {\bf SB08}  is LB19. The infrared counterpart, LB19a, is detected at a separation of 1.7".

The source {\bf SB09} does not have any counterpart in the near-IR, Spitzer or WISE data. There is a detection with AKARI/FIS at a separation of 19" SW (namely, 0531219+120513, see SB09 panel in Figure~\ref{saboca_fcs}), however  this AKARI source is located at $\sim$5" from the bright WISE source WISE~J053121.99+120507.9 
(LB25c in BGH16), so they are probably associated. Therefore, SB09 remains as a probable starless core.

For {\bf SB11} we detected two possible counterparts in the Spitzer/IRAC images. One of them is detected at 3.5 and 4.5\,$\mu$m  at a 
separation of 3.6".  The second one is very faint and only detected at the IRAC I4 channel at 8.0 $\mu$m at a separation of 3.1".  
With this information it is difficult to assign the correct counterpart. There is one bright mid-IR source detected (with WISE too), but at 7.5" from the source.

The {\bf SB12}  source does not have any counterpart in the near-IR, Spitzer or WISE data. There is, however, a detection in the Planck Catalog of compact sources, G192.65-11.70, at a separation of  14.2" from SB12, and at 5.5" from the APEX/LABOCA source B30-LB27, classified as an extended source. The fact that the latter is not detected with ALMA points towards an extended clump (the angular size measured with APEX/LABOCA is 29") and its nature (transient or not) is unclear.

For {\bf SB13} there is only one counterpart detected at 4.4" in the near-IR. The source is marginally seen at 3.6\,$\mu$m.
The fact that is not detected at longer IRAC wavelengths suggests that is probably a contaminant.

Only one counterpart is detected for {\bf SB14}  in the 8\,$\mu$m IRAC image at 4.2" 
separation. There is a source detected with WISE that seems to be related to the IRAC detection, although with SNR between 3 and 10
at 3.4$\mu$m and 12$\mu$m, and upper limits in the other two bands (WISE J053111.14+120624.4).   If the infrared source is related with SB12, the SED is 
consistent with a young object, although additional data is needed for its characterization.

For {\bf SB15} there is an IR counterpart located 5" from the SABOCA detection. The SED only includes four data points with a very uncertain 
3.5\,$\mu$m measurement. We cannot conclude if it is related to the SABOCA source.

Only one infrared counterpart was found for {\bf SB16} at 4.6" (LB28a in BGH16). It is classified as a member, but its SED (with points up to 4.5$\mu$m) does not show a clear infrared excess, therefore the IR and sub-mm sources are probably unrelated.

Finally, we note  that, in addition to SB06, we did not find IR counterparts to the SABOCA sources  
SB01, SB09 (B30-LB20), SB10, SB12 (B30-LB27), and SB17.  In the case of SB06, SB09 and SB12,  they are associated with LABOCA cores, however we verified
that they were not included  in the field of view of the  ALMA images. This is not the  case for SB03, which is included in one of the ALMA
pointings, but is undetected. Therefore,  we remain with six SABOCA sources that are good candidates to be starless cores.
Higher quality observations will help us to understand the true nature of these sources, since
our SABOCA map is not sensitive enough to make a detailed study.


\section{Discussion\label{sec:discussion}}

The ALMA  data have allowed us to characterize two very young stellar members of B30 through the analysis of
their continuum and gas emission.  We have also detected three sources without IR counterparts.

In the following subsections, we discuss the low rate of ALMA detections in the sample of observed APEX/LABOCA cores, 
and compare the properties of the cores with and without ALMA detections.
Finally, we also discuss the possible nature of the three ALMA detections without IR counterparts.

\subsection{ALMA detection rate}

We observed a total of 30 cores previously detected with APEX/LABOCA and with flux densities 
between $\sim$30 and $\sim$190\,mJy, with SNR $\geq3$, and mainly spatially unresolved (see Table~2 from BGH16). 
Since the APEX/LABOCA beam is $\sim$27", we estimate sizes smaller than $\sim$10800\,AU in diameter 
for the unresolved cores.

In the ALMA observations presented here we have only detected  5 out of 30 sources, corresponding to a 
detection rate of 17\%. This can be related to our sensitivity or/and to the size of the emitting sources. The average $rms$ of our ALMA observations is $\sim$0.13\,mJy. In the case of the faintest APEX/LABOCA source, with a flux density of $\sim$29\,mJy, we would expect a flux per beam of 0.8\,mJy if we assume the less favourable case in which the size of the source is as large as the  size given by the ALMA LAS ($\sim$6"). Even in this unfavourable case, ALMA could have detected the APEX/LABOCA sources with SNR$\ge$6.  Since we do not detect most of the APEX/LABOCA cores, we conclude that the low rate of ALMA detections is related to the size of the emitting sources.  Based on the LAS, we estimate that most of the APEX/LABOCA cores 
have sizes larger than 2400\,AU (6" at 400\,pc), or 1200\,AU in radius, as also suggested  by the fact that in all but one of the ALMA detections (LB14)  the flux missed by ALMA is extremely large ($>$85\%, see Table~1). These APEX/LABOCA cores, which remained undetected by ALMA and seem to be dominated by large-scale emission, may be transient cores.


\begin{table*}
\caption{Estimated critical densities (at 15~K) and maximum radii for the three ALMA pre-BD core candidates to be gravitationally unstable.}\label{gravun}
\begin{tabular}{lll lll lll ll} 
\hline 
& \multicolumn{5}{c}{APEX/LABOCA} 
& \multicolumn{5}{c}{ALMA} 
\\    
& $n_{\rm crit}\,^{\rm a}$ 
& $n_{\rm obs}\,^{\rm b}$ 
& $R_{\rm max}\,^{\rm c}$
& $R_{\rm obs}\,^{\rm d}$
&dynamical
& $n_{\rm crit}\,^{\rm a}$ 
& $n_{\rm obs}\,^{\rm b}$ 
& $R_{\rm max}\,^{\rm c}$
& $R_{\rm obs}\,^{\rm d}$ 
& dynamical \\
Core
& [cm$^{-3}$]
& [cm$^{-3}$] 
& [AU]        
& [AU] 
& state
& [cm$^{-3}$]
& [cm$^{-3}$] 
& [AU]        
& [AU] 
& state
\\ \hline
LB08  & 3.3$\times10^6$         & 2.5$\times10^4$       & 971 &  5400            &stable/transient       & 5.0$\times10^8$       & >1.3$\times10^8$      &82      &  <140    &infalling?    \\
LB10  & 1.7$\times10^7$         & >1.0$\times10^4$      & 421 & 1200$^*$--5520         &stable?                        & 3.4$\times10^9$       & >4.9$\times10^7$      &28     &<140         &infalling?\\
LB31  & 5.5$\times10^6$         & >1.8$\times10^4$      & 751 & 1200$^*$--5520  &stable?                        & 4.5$\times10^{10}$& >1.3$\times10^7$    &8              &<140           &infalling?\\  \hline
\end{tabular}
\tablefoot{ 
\tablefoottext{\rm a}{Critical density estimated following equation (3), and using the corresponding APEX/LABOCA or ALMA mass for a temperature of 15~K, as given in Table~1; }  
\tablefoottext{\rm b}{Measured density using the APEX/LABOCA or ALMA masses given in Table~1, obtained for a temperature of 15~K, and using the measured size or the beam size as upper limit if the source is unresolved; }
\tablefoottext{\rm c}{ Maximum radii corresponding to the critical density given in this table, and using the mass used to obtain the critical density; }
\tablefoottext{\rm d}{Measured radii in the APEX/LABOCA or ALMA observations. If the sources are unresolved we give half of the beam size as un upper limit for the radius. The lower limit of the radius in the APEX/LABOCA column corresponds to half the LAS. We adopted this value as a lower limit for the radius of the APEX/LABOCA core given that >90\% of the flux is filtered out by ALMA.}
}
\end{table*}

{\bf 
\subsection{Comparison of LABOCA cores with and without ALMA detections}}

We compared the properties of the LABOCA cores with and without ALMA detections.
First of all, we compared the spatial distribution of the five ALMA sources within the B30 cloud.
They are located both in the northern parts of the cloud (LB08, LB10 and LB14) and in the ionization front (LB19 and LB31).
This is also the case for the 25 LABOCA sources undetected with ALMA (see Figure~\ref{fig:wideALMA}), 
so we do not see any  relation between the detected sources and their position within the cloud.

Regarding the masses,  
the five ALMA detections are associated with APEX/LABOCA cores with masses between 0.046 and 0.182\,M $_{\odot}$, comparable with the undetected cores with masses between 0.043 and 0.278\,M$_{\odot}$.

Finally, we compared the estimated sizes: from the 30 cores observed with ALMA, only eight were spatially resolved with APEX/LABOCA, with sizes (diameters) between  27"~and 62" (this is the deconvolved size after fitting  a Gaussian to the cores, see BGH16). In the case of ALMA, we detected five 
point-like sources, only one being spatially resolved in the LABOCA data: LB08 with 27". 

Therefore, we do not report any particular property of  the ALMA detected cores when compared with the non-detected ones.
As explained in the previous subsection, the main difference should be that the APEX/LABOCA cores undetected with ALMA
have most of their mass in large scales (scales larger than the LAS of ALMA observations).

\subsection{Nature of ALMA detections without infrared counterparts}\label{alma_prebd}

If we assume that LB08, LB10 and LB31 are associated with B30, and since they do not show IR counterparts up to 24$\mu$m (in the case of LB08, up to 70$\mu$m), they could
 be classified either as deeply  embedded protostars undetected in the Spitzer observations \citep[see e.g.,][]{Stutz2013}, or as starless cores.
 
 In the protostar scenario, we can provide  a limit to the luminosity for the central sources.
 LB08 is undetected in the MIPS 70$\mu$m image, however
we have  estimated a background intensity at the ALMA position of $\sim$0.22\,Jy/beam.
If LB08 had a deeply embedded and undetected counterpart at this
wavelength, we could use this flux  as an upper limit to the emission of a source at the given position.
We could then estimate an upper limit to the internal luminosity of a potential counterpart using the 
formula by \citet{Dunham2008}, scaled to the distance of B30. By doing this, we derived a value of L$_{int}$$<$0.1\,L$_{\odot}$, which is the luminosity 
limit established for VeLLOs. As a different approach, we can use the limiting magnitudes at all wavelengths (they have been estimated in BGH16, see their Table 1), together with the APEX/LABOCA fluxes (instead of the ALMA fluxes that can suffer significant levels of spatial filtering when compared with single-dish observations), to integrate the SED of the object (see Figure~\ref{alma_sed}) and  derive upper limits to the bolometric luminosity and bolometric temperature. We 
estimate L$_{bol}$ $<$0.08\,L$_{\odot}$  and  T$_{bol}$ $>$ 54\,K, respectively.
These values suggest that the source could be a VeLLO, with a lower limit of the T$_{bol}$ consistent with a Class~0 object 
\citep[Class~0 sources show T$_{bol}$ $<$ 90\,K,][]{Chen1995} or more evolved.
The value of L$_{bol}$ $<$0.07\,L$_{\odot}$ implies that the internal luminosity should be even smaller.

In the case of LB10 and LB31, there are no 70\,$\mu$m data available at their positions. Since LB31 has an upper limit at 350$\mu$m
 we can estimate L$_{bol}$ $<$0.05\,L$_{\odot}$  and  T$_{bol}$ $>$ 21\,K 
by integrating its SED (displayed in Figure~\ref{alma_sed}). This means that LB31 could also be a VeLLO with a T$_{bol}$ consistent with a Class~0 (or more evolved) object, although this is a very rough estimation due to the lack of data-points in the sub-mm range.  The fact that this source is well within the ionization front of the B30 cloud (see Figure~1) suggests that it could be a photo-evaporated proto-BD. In the case of LB10, there is no data available between 24$\mu$m and 870$\mu$m. 
Hence,  it is difficult to estimate a limit for the luminosity for this object, since the sub-mm range is not well-sampled. 

A second possibility that we explored is that the three ALMA sources are starless cores.
For a $T_{\rm d}$ of 15\,K, their  estimated masses  are between 0.9 and 9\,M$_{\rm Jup}$ using the ALMA data, and  between $\sim$46 and 106\,M$_{\rm Jup}$ using the  APEX/LABOCA data. Of special interest are the cases of LB10 and LB31, with APEX/LABOCA masses below (46\,M$_{\rm Jup}$) or very close to the substellar regime (82\,M$_{\rm Jup}$), implying that they could be pre-substellar cores.  

If these three cores collapse to form stars, we can estimate the final masses of the central objects 
assuming  that the core formation efficiency in low-mass dense cores is $\sim$30\%  \citep[e.g.,][]{Motte1998,Alves2007,Bontemps2010}.
If we add 30\% of the LABOCA mass to the ALMA masses, we obtain  $\sim$41\,M$_{\rm Jup}$ (LB08),   $\sim$17\,M$_{\rm Jup}$ (LB10), and $\sim$26\,M$_{\rm Jup}$ (LB31),  all of them below the sub-stellar mass regime. Therefore, the three ALMA sources  could be considered as pre-BD core candidates.
We note that the derived masses are strongly dependent on the dust temperature. A colder temperature, such as 10\,K, would increase the final masses by a factor of approximately two, which would also be consistent with pre-BD core candidates in LB10 and LB31. LB08 would  be at the border between a low-mass prestellar object and a pre-BD core.

 In the framework of the turbulent fragmentation theory, \citet{Padoan2004} have explained that 
  substellar objects can form in cores with a density as high as the critical density for the collapse of a  
  BD mass core. To study if  the three LABOCA starless cores detected with ALMA might form BDs,  we 
  followed the same procedure as \citet{deGregorio2016}, and compared the density of the LABOCA cores with  the critical 
  density ($n_{\rm crit}$) of a Bonnor-Ebert (BE) isothermal sphere, following the relation:

\begin{equation}\label{BE}
\centering
\Bigg[\frac{n_{\rm crit}}{\rm{cm}^{-3}}\Bigg] = 1.089\times10^4\;\Bigg[\frac{M_{\rm BE}}{M_\odot}\Bigg]^{-2}\;\Bigg[\frac{T}{10~\rm K}\Bigg]^3,
\end{equation}

\noindent
where $M_{\rm BE}$ is the mass of the BE sphere.
For these three cores, we have first assumed that the mass of the BE sphere is the one obtained with LABOCA for a typical 
temperature  of 15~K (Table~1), and calculated $n_{\rm crit}$ following equation (3). We also estimated, given the mass 
of the object, the radius $R_{\rm max}$ corresponding to $n_{\rm crit}$, and compared it to the measured radius (only LB08 
is resolved, for the other cores we took the beam size as an upper limit). The results are shown in Table~3.

Table~3 shows that, if we take into account only APEX/LABOCA data, LB08 is not gravitationally unstable, because its density is smaller than the critical density by approximately two orders of magnitude, and its radius is larger than the maximum radius required for instability. For the case of LB10 and LB31, this is not so clear, because they are not resolved by APEX and the observed densities are only lower limits. However, if we compare $R_{\rm max}$ with the possible sizes of the cores, we again find that also LB10 and LB31 should not be gravitationally unstable. This is because, although the cores are not resolved, ALMA has filtered out even more flux than for the case of LB08 (i.e., >90\%, see Table~1), strongly suggesting that most of the mass measured by APEX/LABOCA comes from spatial scales much larger than the ALMA LAS, that is, much larger than 1200~AU, and probably  larger than $R_{\rm max}$, especially for the case of LB10. Thus, using only the APEX/LABOCA data, it would rather seem, following the \citet{Padoan2004} theory, that the three cores are not gravitationally unstable.

This poses a major problem, because ALMA has revealed that there already seems to be a very compact object at the center of the APEX/LABOCA cores, indicative of on-going collapse. 
In fact, the calculated upper limits to the observed densities using ALMA remain consistent with infalling objects (mainly in the case of LB08, which is of the same order as the estimated critical density). 
In addition, very recent studies  conclude that a shock condition for the turbulence fragmentation scenario to be efficient is a very unlikely configuration of converging flows \citep{Lomax2016}, again posing serious problems to the turbulent fragmentation theory for BD formation.

An alternative scenario is to assume that the three ALMA  pre-BD core candidates are already the result of gravitational contraction,
constituting the `tip of the iceberg' of a larger-scale collapse. In this case, the collapse is expected to occur from the outside-in
and to self-consistently develop a near $r^{-2}$ density profile but with a finite infall velocity during their prestellar evolution  
\citep[e.g.,][]{Larson1969,Whitworth1985, Gomez2007,Gong2009,Mohammadpour2013, Naranjo2015}. 
If the cores have indeed an $r^{-2}$ profile, and we assume that the masses inside the APEX beam are the ones given in Table~1,  
we can estimate the expected masses inside the ALMA beam (since we measure $\sim$0.9"$\times$0.5", 
we have assumed an average value of 0.7 arcsec, corresponding to 140\,AU of radius). 
The result is displayed in Table~\ref{r2profile}. The estimated masses are $\sim$3, 1, and 2\,M$_{\rm Jup}$ for LB08, LB10 and LB31, respectively. Interestingly, they are consistent with the measured ones within the uncertainties (typically of a factor of 4), suggesting that 
  the density profile is indeed not far from $r^{-2}$, and supporting the 
  hypothesis that the whole core is already the result of a gravitational 
  collapse. Further observations aimed at studying the kinematics of the three ALMA sources would help to elucidate whether or not their hosting cores are indeed infalling towards the center. These observations would also help to estimate the final mass of the future collapsed objects and thus assess (or not)  their pre-BD nature.

Finally, we stress that with our current data set we cannot exclude the possibility that these three sources are extragalactic contaminants not included in the NED database.  Although their location in the cloud (well within the filaments, at least in the case of LB08 and LB31) suggests that they are probably associated to B30, future ALMA observations will allow confirmation of their B30 membership.

\begin{table}
\caption{Estimated masses (at 15~K) of the three ALMA pre-BD core candidates  within an ALMA beam assuming a $r^{-2}$ density profile.}\label{r2profile}
\begin{tabular}{lllll} 
\hline 
& \multicolumn{2}{c}{APEX/LABOCA} 
& \multicolumn{2}{c}{ALMA Mass} 
\\    
& Mass 
& Radius  
& Estimated $^*$
& Observed \\
Core
& [M$_{\rm Jup}$] 
& [AU]        
& [M$_{\rm Jup}$] 
& [M$_{\rm Jup}$] 
\\ \hline
LB08  & 106 & 5400   &  3 & 9  \\
LB10  & 46   & 5520   &  1 & 3   \\
LB31  & 82  & 5520   &  2 & 0.9 \\ \hline
\end{tabular}

$^*$ Estimated for an average beam size of 0.7", or 140\,AU of radius.
\end{table}

\section{Conclusions\label{sec:conclusions}}

We analyzed ALMA  observations of the young dark cloud B30 at 880\,$\mu$m. 
We complemented the ALMA observations with APEX/SABOCA data obtained and 350\,$\mu$m  
and covering the southern part of the B30 cloud.  The aim of our study was to  shed light on the nature of 30 
sub-mm sources previously detected with APEX/LABOCA at 870\,$\mu$m.
Our main results can be summarized as follows:

   \begin{itemize}
      
 \item We have detected 5 (out of 30) compact sources with ALMA with masses between 0.9 and 67\,$M_{\rm Jup}$. 
 Two of them show clear infrared counterparts (LB14 and LB19),  while three of them do not (LB08, LB10 and LB31).  
 
 \item The properties of  LB14 and LB19 are consistent with a Class~II and a Class~I low-mass stellar object, respectively. 
 Both sources show gas emission that allows for their confirmation as B30 members. LB14 shows very compact gas emission compatible with a rotating disk.
 LB19 shows two gas emission components: a compact one, with an asymmetric spectrum in  which the 
 blueshifted emission is stronger than the red one, and an extended component of $\sim$2400\,AU surrounding the compact emission. 
 The kinematical analysis of the two components suggests that we might be observing a rotating oblate envelope, 
 and infalling signatures at 400\,AU scales. 
 
\item { If LB08, LB10 and LB31 belong to B30, they can be classified as either
 deeply embedded protostars, or starless cores. In the former scenario, two of the sources (LB08 and LB31) show internal luminosity upper limits 
consistent with VeLLOs, while LB10 does not have enough data to  estimate a reliable limit.
 In the latter scenario,} the estimated final masses of the three sources are near the substellar regime. Therefore,
 they could be pre-BD core candidates if B30 members. One of the sources (LB31) is located at the ionization front of the B30 
 cloud and might be a photo-evaporated  proto-BD or pre-BD core candidate.
 
  \item According to the theory of BD formation through turbulent
    fragmentation, the three starless LABOCA cores with ALMA
    detections are more consistent with gravitationally stable
    cores. However, this result is not consistent with the fact that
    ALMA detected a compact source at the center of these cores.  As
    an alternate scenario, we propose that these cores are the result
    of a large scale gravitational contraction.  In this case, the
    estimated masses using a $r^{-2}$ density profile (which is
    characteristic of a collapsing core) are consistent with the
    observed ones within the uncertainties.
 
      \item We detected 17 sources with SABOCA at 350\,$\mu$m in the
        southern part of the cloud{ but with very low SNR (between 3.0
          and 4.6)}. Eleven show infrared counterparts, although in
        most of the cases they are detected at separation larger than
        4" from the center of the SABOCA beam, so their association is
        uncertain.  Six objects do not show IR counterparts and may be
        considered starless cores.



   \end{itemize}

Future observations will allow confirmation of the true nature of the three ALMA sources without IR counterparts.

\begin{acknowledgements}
 
 This project has been funded by the BBVA Foundation under the {\it
   Convocatoria 2015 de Ayudas Fundaci\'on BBVA a Investigadores y
   Creadores Culturales}.  We thank E. V\'azquez-Semadeni and J.~Alves
 for very useful comments, and R.~Lorente and I.Yamamura for their
 help with the AKARI data.  We are indebted to the the Calar Alto
 Observatory staff for their excellent work taking the near-IR data
 under the Service Mode program.  NH thanks the ALMA Science Center in
 Santiago for hosting her during 2.5 months.  IdG acknowledges support
 fromMICINN (Spain) AYA2011-30228-C03 grant (including FEDER funds).
 AP and LZ acknowledge the financial support from UNAM-DGAPA-PAPITT
 IA102815 grant (M\'exico).  DB has been funded by Spanish grant
 AYA2012-38897-C02-01.  HB is funded by the the Ram\'on y Cajal
 fellowship program number RYC-2009-04497.  AB acknowledges financial
 support from the Proyecto Fondecyt de Iniciaci\'on 11140572, and
 scientific support from the Millenium Science Initiative, Chilean
 Ministry of Economy, Nucleus RC130007.  MTR acknowledges partial
 support from CATA (PB06, CONICYT-Chile). C.E. is partly supported by
 Spanish Grant AYA 2014-55840-P.  This publication makes use of data
 products from the Wide-field Infrared Survey Explorer, which is a
 joint project of the University of California, Los Angeles, and the
 Jet Propulsion Laboratory/California Institute of Technology, funded
 by the National Aeronautics and Space Administration.
\end{acknowledgements}        

%
\bibliographystyle{aa} 
\bibliography{almaB30}

\begin{thebibliography}{73}
\expandafter\ifx\csname natexlab\endcsname\relax\def\natexlab#1{#1}\fi

\bibitem[{{Allard} {et~al.}(2012){Allard}, {Homeier}, {Freytag}, \&
  {Sharp}}]{Allard2012.1}
{Allard}, F., {Homeier}, D., {Freytag}, B., \& {Sharp}, C.~M. 2012, in EAS
  Publications Series, Vol.~57, EAS Publications Series, ed. C.~{Reyl{\'e}},
  C.~{Charbonnel}, \& M.~{Schultheis}, 3--43

\bibitem[{{Allen} {et~al.}(2004){Allen}, {Calvet}, {D'Alessio}, {Merin},
  {Hartmann}, {Megeath}, {Gutermuth}, {Muzerolle}, {Pipher}, {Myers}, \&
  {Fazio}}]{Allen2004}
{Allen}, L.~E., {Calvet}, N., {D'Alessio}, P., {et~al.} 2004, \apjs, 154, 363

\bibitem[{{Alves} {et~al.}(2007){Alves}, {Lombardi}, \& {Lada}}]{Alves2007}
{Alves}, J., {Lombardi}, M., \& {Lada}, C.~J. 2007, \aap, 462, L17

\bibitem[{{Alves de Oliveira} {et~al.}(2012){Alves de Oliveira}, {Moraux},
  {Bouvier}, \& {Bouy}}]{Alves2012}
{Alves de Oliveira}, C., {Moraux}, E., {Bouvier}, J., \& {Bouy}, H. 2012, \aap,
  539, A151

\bibitem[{{Andr{\'e}} {et~al.}(2012){Andr{\'e}}, {Ward-Thompson}, \&
  {Greaves}}]{Andre2012}
{Andr{\'e}}, P., {Ward-Thompson}, D., \& {Greaves}, J. 2012, Science, 337, 69

\bibitem[{{Barrado} {et~al.}(2016){Barrado}, {de Gregorio-Monsalvo}, {Huelamo},
  {Morales-Calderon}, {Bayo}, {Palau}, {Ruiz}, {Riviere-Marichalar}, {Bouy},
  {Morata}, {Stauffer}, \& {Eiroa}}]{Barrado2016}
{Barrado}, D., {de Gregorio-Monsalvo}, I., {Huelamo}, N., {et~al.} 2016, \aap,
  submitted

\bibitem[{{Barrado} {et~al.}(2009){Barrado}, {Morales-Calder{\'o}n}, {Palau},
  {Bayo}, {de Gregorio-Monsalvo}, {Eiroa}, {Hu{\'e}lamo}, {Bouy}, {Morata}, \&
  {Schmidtobreick}}]{Barrado2009}
{Barrado}, D., {Morales-Calder{\'o}n}, M., {Palau}, A., {et~al.} 2009, \aap,
  508, 859

\bibitem[{{Barrado y Navascu{\'e}s} \& {Jayawardhana}(2004)}]{Barrado2004.1}
{Barrado y Navascu{\'e}s}, D. \& {Jayawardhana}, R. 2004, \apj, 615, 840

\bibitem[{{Barrado y Navascu{\'e}s} {et~al.}(2004){Barrado y Navascu{\'e}s},
  {Stauffer}, {Bouvier}, {Jayawardhana}, \& {Cuillandre}}]{Barrado2004.3}
{Barrado y Navascu{\'e}s}, D., {Stauffer}, J.~R., {Bouvier}, J.,
  {Jayawardhana}, R., \& {Cuillandre}, J.-C. 2004, \apj, 610, 1064

\bibitem[{{Barrado y Navascu{\'e}s} {et~al.}(2007){Barrado y Navascu{\'e}s},
  {Stauffer}, {Morales-Calder{\'o}n}, {Bayo}, {Fazzio}, {Megeath}, {Allen},
  {Hartmann}, \& {Calvet}}]{Barrado2007.1}
{Barrado y Navascu{\'e}s}, D., {Stauffer}, J.~R., {Morales-Calder{\'o}n}, M.,
  {et~al.} 2007, \apj, 664, 481

\bibitem[{{Bate}(2012)}]{Bate2012}
{Bate}, M.~R. 2012, \mnras, 419, 3115

\bibitem[{{Bate} {et~al.}(2002){Bate}, {Bonnell}, \& {Bromm}}]{Bate2002}
{Bate}, M.~R., {Bonnell}, I.~A., \& {Bromm}, V. 2002, \mnras, 332, L65

\bibitem[{{Bayo}(2009)}]{Bayo2009.1}
{Bayo}, A. 2009, PhD dissertations, Universidad Aut\'onoma de Madrid

\bibitem[{{Bayo} {et~al.}(2011){Bayo}, {Barrado}, {Stauffer},
  {Morales-Calder{\'o}n}, {Melo}, {Hu{\'e}lamo}, {Bouy}, {Stelzer}, {Tamura},
  \& {Jayawardhana}}]{Bayo2011.1}
{Bayo}, A., {Barrado}, D., {Stauffer}, J., {et~al.} 2011, \aap, 536, A63

\bibitem[{{Bonnell} {et~al.}(2008){Bonnell}, {Clark}, \& {Bate}}]{Bonnell2008}
{Bonnell}, I.~A., {Clark}, P., \& {Bate}, M.~R. 2008, \mnras, 389, 1556

\bibitem[{{Bontemps} {et~al.}(2010){Bontemps}, {Motte}, {Csengeri}, \&
  {Schneider}}]{Bontemps2010}
{Bontemps}, S., {Motte}, F., {Csengeri}, T., \& {Schneider}, N. 2010, \aap,
  524, A18

\bibitem[{{Bourke} {et~al.}(2006){Bourke}, {Myers}, {Evans}, {Dunham},
  {Kauffmann}, {Shirley}, {Crapsi}, {Young}, {Huard}, {Brooke}, {Chapman},
  {Cieza}, {Lee}, {Teuben}, \& {Wahhaj}}]{Bourke2006}
{Bourke}, T.~L., {Myers}, P.~C., {Evans}, II, N.~J., {et~al.} 2006, \apjl, 649,
  L37

\bibitem[{{Bouy} {et~al.}(2007){Bouy}, {Hu{\'e}lamo}, {Mart{\'{\i}}n}, {Barrado
  Y Navascu{\'e}s}, {Sterzik}, \& {Pantin}}]{Bouy2007}
{Bouy}, H., {Hu{\'e}lamo}, N., {Mart{\'{\i}}n}, E.~L., {et~al.} 2007, \aap,
  463, 641

\bibitem[{{Bouy} {et~al.}(2009){Bouy}, {Hu{\'e}lamo}, {Mart{\'{\i}}n},
  {Marchis}, {Barrado Y Navascu{\'e}s}, {Kolb}, {Marchetti}, {Petr-Gotzens},
  {Sterzik}, {Ivanov}, {K{\"o}hler}, \& {N{\"u}rnberger}}]{Bouy2009}
{Bouy}, H., {Hu{\'e}lamo}, N., {Mart{\'{\i}}n}, E.~L., {et~al.} 2009, \aap,
  493, 931

\bibitem[{{Caballero} {et~al.}(2007){Caballero}, {B{\'e}jar}, {Rebolo},
  {Eisl{\"o}ffel}, {Zapatero Osorio}, {Mundt}, {Barrado Y Navascu{\'e}s},
  {Bihain}, {Bailer-Jones}, {Forveille}, \& {Mart{\'{\i}}n}}]{Caballero2007}
{Caballero}, J.~A., {B{\'e}jar}, V.~J.~S., {Rebolo}, R., {et~al.} 2007, \aap,
  470, 903

\bibitem[{{Cambresy} {et~al.}(1997){Cambresy}, {Epchtein}, {Copet}, {de Batz},
  {Kimeswenger}, {Le Bertre}, {Rouan}, \& {Tiphene}}]{Cambresy1997}
{Cambresy}, L., {Epchtein}, N., {Copet}, E., {et~al.} 1997, \aap, 324, L5

\bibitem[{{Chen} {et~al.}(1995){Chen}, {Myers}, {Ladd}, \& {Wood}}]{Chen1995}
{Chen}, H., {Myers}, P.~C., {Ladd}, E.~F., \& {Wood}, D.~O.~S. 1995, \apj, 445,
  377

\bibitem[{{Connelley} {et~al.}(2008){Connelley}, {Reipurth}, \&
  {Tokunaga}}]{Connelley2008}
{Connelley}, M.~S., {Reipurth}, B., \& {Tokunaga}, A.~T. 2008, \aj, 135, 2496

\bibitem[{{de Gregorio-Monsalvo} {et~al.}(2016){de Gregorio-Monsalvo},
  {Barrado}, {Bouy}, {Bayo}, {Palau}, {Morales-Calder{\'o}n}, {Hu{\'e}lamo},
  {Morata}, {Mer{\'{\i}}n}, \& {Eiroa}}]{deGregorio2016}
{de Gregorio-Monsalvo}, I., {Barrado}, D., {Bouy}, H., {et~al.} 2016, \aap,
  590, A79

\bibitem[{{Delfosse} {et~al.}(1997){Delfosse}, {Tinney}, {Forveille},
  {Epchtein}, {Bertin}, {Borsenberger}, {Copet}, {de Batz}, {Fouque},
  {Kimeswenger}, {Le Bertre}, {Lacombe}, {Rouan}, \& {Tiphene}}]{Delfosse1997}
{Delfosse}, X., {Tinney}, C.~G., {Forveille}, T., {et~al.} 1997, \aap, 327, L25

\bibitem[{{Delorme} {et~al.}(2008){Delorme}, {Willott}, {Forveille},
  {Delfosse}, {Reyl{\'e}}, {Bertin}, {Albert}, {Artigau}, {Robin}, {Allard},
  {Doyon}, \& {Hill}}]{Delorme2008}
{Delorme}, P., {Willott}, C.~J., {Forveille}, T., {et~al.} 2008, \aap, 484, 469

\bibitem[{{di Francesco} {et~al.}(2007){di Francesco}, {Evans}, {Caselli},
  {Myers}, {Shirley}, {Aikawa}, \& {Tafalla}}]{diFrancesco2007}
{di Francesco}, J., {Evans}, II, N.~J., {Caselli}, P., {et~al.} 2007,
  Protostars and Planets V, 17

\bibitem[{{Dolan} \& {Mathieu}(1999)}]{Dolan1999.1}
{Dolan}, C.~J. \& {Mathieu}, R.~D. 1999, \aj, 118, 2409

\bibitem[{{Dolan} \& {Mathieu}(2001)}]{Dolan2001.1}
{Dolan}, C.~J. \& {Mathieu}, R.~D. 2001, \aj, 121, 2124

\bibitem[{{Dolan} \& {Mathieu}(2002)}]{Dolan2002.1}
{Dolan}, C.~J. \& {Mathieu}, R.~D. 2002, \aj, 123, 387

\bibitem[{{Duerr} {et~al.}(1982){Duerr}, {Imhoff}, \& {Lada}}]{Duerr82}
{Duerr}, R., {Imhoff}, C.~L., \& {Lada}, C.~J. 1982, \apj, 261, 135

\bibitem[{{Dunham} {et~al.}(2008){Dunham}, {Crapsi}, {Evans}, {Bourke},
  {Huard}, {Myers}, \& {Kauffmann}}]{Dunham2008}
{Dunham}, M.~M., {Crapsi}, A., {Evans}, II, N.~J., {et~al.} 2008, \apjs, 179,
  249

\bibitem[{{Fitzpatrick}(1999)}]{Fitzpatrick1999}
{Fitzpatrick}, E.~L. 1999, \pasp, 111, 63

\bibitem[{{Forbrich} {et~al.}(2015){Forbrich}, {Lada}, {Lombardi},
  {Rom{\'a}n-Z{\'u}{\~n}iga}, \& {Alves}}]{Forbrich2015}
{Forbrich}, J., {Lada}, C.~J., {Lombardi}, M., {Rom{\'a}n-Z{\'u}{\~n}iga}, C.,
  \& {Alves}, J. 2015, \aap, 580, A114

\bibitem[{{G{\'o}mez} {et~al.}(2007){G{\'o}mez}, {V{\'a}zquez-Semadeni},
  {Shadmehri}, \& {Ballesteros-Paredes}}]{Gomez2007}
{G{\'o}mez}, G.~C., {V{\'a}zquez-Semadeni}, E., {Shadmehri}, M., \&
  {Ballesteros-Paredes}, J. 2007, \apj, 669, 1042

\bibitem[{{Gomez} \& {Lada}(1998)}]{Gomez98}
{Gomez}, M. \& {Lada}, C.~J. 1998, \aj, 116, 1508

\bibitem[{{Gong} \& {Ostriker}(2009)}]{Gong2009}
{Gong}, H. \& {Ostriker}, E.~C. 2009, \apj, 699, 230

\bibitem[{{Hennebelle} \& {Chabrier}(2008)}]{Hennebelle2008}
{Hennebelle}, P. \& {Chabrier}, G. 2008, \apj, 684, 395

\bibitem[{{Hester} {et~al.}(1996){Hester}, {Scowen}, {Sankrit}, {Lauer},
  {Ajhar}, {Baum}, {Code}, {Currie}, {Danielson}, {Ewald}, {Faber},
  {Grillmair}, {Groth}, {Holtzman}, {Hunter}, {Kristian}, {Light}, {Lynds},
  {Monet}, {O'Neil}, {Shaya}, {Seidelmann}, \& {Westphal}}]{Hester1996}
{Hester}, J.~J., {Scowen}, P.~A., {Sankrit}, R., {et~al.} 1996, \aj, 111, 2349

\bibitem[{{Hodapp} {et~al.}(2009){Hodapp}, {Iserlohe}, {Stecklum}, \&
  {Krabbe}}]{Hodapp2009}
{Hodapp}, K.~W., {Iserlohe}, C., {Stecklum}, B., \& {Krabbe}, A. 2009, \apjl,
  701, L100

\bibitem[{{Kov{\'a}cs}(2008)}]{Kovacs2008}
{Kov{\'a}cs}, A. 2008, in \procspie, Vol. 7020, Millimeter and Submillimeter
  Detectors and Instrumentation for Astronomy IV, 70201S

\bibitem[{{Lang} {et~al.}(2000){Lang}, {Masheder}, {Dame}, \&
  {Thaddeus}}]{Lang2000}
{Lang}, W.~J., {Masheder}, M.~R.~W., {Dame}, T.~M., \& {Thaddeus}, P. 2000,
  \aap, 357, 1001

\bibitem[{{Larson}(1969)}]{Larson1969}
{Larson}, R.~B. 1969, \mnras, 145, 271

\bibitem[{{Lee} {et~al.}(2009){Lee}, {Bourke}, {Myers}, {Dunham}, {Evans},
  {Lee}, {Huard}, {Wu}, {Gutermuth}, {Kim}, \& {Kang}}]{Lee2009}
{Lee}, C.~W., {Bourke}, T.~L., {Myers}, P.~C., {et~al.} 2009, \apj, 693, 1290

\bibitem[{{Lee} {et~al.}(2013){Lee}, {Kim}, {Kim}, {Saito}, {Myers}, \&
  {Kurono}}]{Lee2013}
{Lee}, C.~W., {Kim}, M.-R., {Kim}, G., {et~al.} 2013, \apj, 777, 50

\bibitem[{{Liu} {et~al.}(2016){Liu}, {Zhang}, {Kim}, {Wu}, {Lee}, {Lee},
  {Tatematsu}, {Choi}, {Juvela}, {Thompson}, {Goldsmith}, {Liu}, {Naomi},
  {Koch}, {Henkel}, {Sanhueza}, {He}, {Rivera-Ingraham}, {Wang}, {Cunningham},
  {Tang}, {Lai}, {Yuan}, {Li}, {Fuller}, {Kang}, {Nguyen Luong}, {Liu},
  {Ristorcelli}, {Yang}, {Xu}, {Hirota}, {Mardones}, {Qin}, {Chen}, {Kwon},
  {Meng}, {Zhang}, {Kim}, \& {Yi}}]{Liu2016}
{Liu}, T., {Zhang}, Q., {Kim}, K.-T., {et~al.} 2016, \apjs, 222, 7

\bibitem[{{Lomax} {et~al.}(2016){Lomax}, {Whitworth}, \& {Hubber}}]{Lomax2016}
{Lomax}, O., {Whitworth}, A.~P., \& {Hubber}, D.~A. 2016, \mnras
  [\eprint[arXiv]{1602.05789}]

\bibitem[{{Mainzer} {et~al.}(2011){Mainzer}, {Cushing}, {Skrutskie}, {Gelino},
  {Kirkpatrick}, {Jarrett}, {Masci}, {Marley}, {Saumon}, {Wright}, {Beaton},
  {Dietrich}, {Eisenhardt}, {Garnavich}, {Kuhn}, {Leisawitz}, {Marsh},
  {McLean}, {Padgett}, \& {Rueff}}]{Mainzer2011}
{Mainzer}, A., {Cushing}, M.~C., {Skrutskie}, M., {et~al.} 2011, \apj, 726, 30

\bibitem[{{Matzner} \& {Levin}(2005)}]{Matzner2005}
{Matzner}, C.~D. \& {Levin}, Y. 2005, \apj, 628, 817

\bibitem[{{Mohammadpour} \& {Stahler}(2013)}]{Mohammadpour2013}
{Mohammadpour}, M. \& {Stahler}, S.~W. 2013, \mnras, 433, 3389

\bibitem[{{Morales-Calder\'on}(2008)}]{Morales2008.1}
{Morales-Calder\'on}, M. 2008, PhD dissertations, Universidad Aut\'onoma de
  Madrid

\bibitem[{{Morata} {et~al.}(2015){Morata}, {Palau}, {Gonz{\'a}lez}, {de
  Gregorio-Monsalvo}, {Ribas}, {Perger}, {Bouy}, {Barrado}, {Eiroa}, {Bayo},
  {Hu{\'e}lamo}, {Morales-Calder{\'o}n}, \& {Rodr{\'{\i}}guez}}]{Morata2015}
{Morata}, O., {Palau}, A., {Gonz{\'a}lez}, R.~F., {et~al.} 2015, \apj, 807, 55

\bibitem[{{Motte} {et~al.}(1998){Motte}, {Andre}, \& {Neri}}]{Motte1998}
{Motte}, F., {Andre}, P., \& {Neri}, R. 1998, \aap, 336, 150

\bibitem[{{Murdin} \& {Penston}(1977)}]{Murdin77.1}
{Murdin}, P. \& {Penston}, M.~V. 1977, \mnras, 181, 657

\bibitem[{{Mu{\v z}i{\'c}} {et~al.}(2015){Mu{\v z}i{\'c}}, {Scholz}, {Geers},
  \& {Jayawardhana}}]{Muzic2015}
{Mu{\v z}i{\'c}}, K., {Scholz}, A., {Geers}, V.~C., \& {Jayawardhana}, R. 2015,
  \apj, 810, 159

\bibitem[{{Naranjo-Romero} {et~al.}(2015){Naranjo-Romero},
  {V{\'a}zquez-Semadeni}, \& {Loughnane}}]{Naranjo2015}
{Naranjo-Romero}, R., {V{\'a}zquez-Semadeni}, E., \& {Loughnane}, R.~M. 2015,
  \apj, 814, 48

\bibitem[{{Ossenkopf} \& {Henning}(1994)}]{Ossenkopf1994}
{Ossenkopf}, V. \& {Henning}, T. 1994, \aap, 291, 943

\bibitem[{{Padoan} \& {Nordlund}(2004)}]{Padoan2004}
{Padoan}, P. \& {Nordlund}, {\AA}. 2004, \apj, 617, 559

\bibitem[{{Palau} {et~al.}(2012){Palau}, {de Gregorio-Monsalvo}, {Morata},
  {Stamatellos}, {Hu{\'e}lamo}, {Eiroa}, {Bayo}, {Morales-Calder{\'o}n},
  {Bouy}, {Ribas}, {Asmus}, \& {Barrado}}]{Palau2012}
{Palau}, A., {de Gregorio-Monsalvo}, I., {Morata}, {\`O}., {et~al.} 2012,
  \mnras, 424, 2778

\bibitem[{{Palau} {et~al.}(2010){Palau}, {S{\'a}nchez-Monge}, {Busquet},
  {Estalella}, {Zhang}, {Ho}, {Beltr{\'a}n}, \& {Beuther}}]{Palau2010}
{Palau}, A., {S{\'a}nchez-Monge}, {\'A}., {Busquet}, G., {et~al.} 2010, \aap,
  510, A5

\bibitem[{{Palau} {et~al.}(2014){Palau}, {Zapata}, {Rodr{\'{\i}}guez}, {Bouy},
  {Barrado}, {Morales-Calder{\'o}n}, {Myers}, {Chapman}, {Ju{\'a}rez}, \&
  {Li}}]{Palau2014}
{Palau}, A., {Zapata}, L.~A., {Rodr{\'{\i}}guez}, L.~F., {et~al.} 2014, \mnras,
  444, 833

\bibitem[{{Perryman} {et~al.}(1997){Perryman}, {Lindegren}, {Kovalevsky},
  {Hoeg}, {Bastian}, {Bernacca}, {Cr{\'e}z{\'e}}, {Donati}, {Grenon}, {van
  Leeuwen}, {van der Marel}, {Mignard}, {Murray}, {Le Poole}, {Schrijver},
  {Turon}, {Arenou}, {Froeschl{\'e}}, \& {Petersen}}]{Perryman1997}
{Perryman}, M.~A.~C., {Lindegren}, L., {Kovalevsky}, J., {et~al.} 1997, \aap,
  323, L49

\bibitem[{{Planck Collaboration} {et~al.}(2014){Planck Collaboration},
  {Abergel}, {Ade}, {Aghanim}, {Alves}, {Aniano}, {Armitage-Caplan}, {Arnaud},
  {Ashdown}, {Atrio-Barandela}, \& et~al.}]{Planck2014}
{Planck Collaboration}, {Abergel}, A., {Ade}, P.~A.~R., {et~al.} 2014, \aap,
  571, A11

\bibitem[{{Reid} {et~al.}(1988){Reid}, {Schneps}, {Moran}, {Gwinn}, {Genzel},
  {Downes}, \& {Roennaeng}}]{Reid1988}
{Reid}, M.~J., {Schneps}, M.~H., {Moran}, J.~M., {et~al.} 1988, \apj, 330, 809

\bibitem[{{Reipurth} \& {Clarke}(2001)}]{Reipurth2001}
{Reipurth}, B. \& {Clarke}, C. 2001, \aj, 122, 432

\bibitem[{{Scholz} {et~al.}(2013){Scholz}, {Geers}, {Clark}, {Jayawardhana}, \&
  {Muzic}}]{Scholz2013}
{Scholz}, A., {Geers}, V., {Clark}, P., {Jayawardhana}, R., \& {Muzic}, K.
  2013, \apj, 775, 138

\bibitem[{{Stutz} {et~al.}(2013){Stutz}, {Tobin}, {Stanke}, {Megeath},
  {Fischer}, {Robitaille}, {Henning}, {Ali}, {di Francesco}, {Furlan},
  {Hartmann}, {Osorio}, {Wilson}, {Allen}, {Krause}, \& {Manoj}}]{Stutz2013}
{Stutz}, A.~M., {Tobin}, J.~J., {Stanke}, T., {et~al.} 2013, \apj, 767, 36

\bibitem[{{White} \& {Basri}(2003)}]{White2003}
{White}, R.~J. \& {Basri}, G. 2003, \apj, 582, 1109

\bibitem[{{Whitworth} \& {Summers}(1985)}]{Whitworth1985}
{Whitworth}, A. \& {Summers}, D. 1985, \mnras, 214, 1

\bibitem[{{Whitworth} \& {Stamatellos}(2006)}]{Whitworth2006}
{Whitworth}, A.~P. \& {Stamatellos}, D. 2006, \aap, 458, 817

\bibitem[{{Whitworth} \& {Zinnecker}(2004)}]{Whitworth2004}
{Whitworth}, A.~P. \& {Zinnecker}, H. 2004, \aap, 427, 299

\bibitem[{{Yamamura} {et~al.}(2009){Yamamura}, {Makiuti}, {Ikeda}, {Fukuda},
  {Yamauchi}, {Hasegawa}, {Nakagawa}, {Narumi}, {Baba}, {Takagi}, {Jeong},
  {Oh}, {Lee}, {Savage}, {Rahman}, {Thomson}, {Oliver}, {Figueredo},
  {Serjeant}, {White}, {Pearson}, {Wang}, {Rowan-Robinson}, {Kester}, {van der
  Wolk}, {Barthel}, {Salama}, {Alfageme}, {Garc{\'{\i}}a-Lario}, {Stephenson},
  {Cohen}, \& {Mueller}}]{Yamamura2009}
{Yamamura}, I., {Makiuti}, S., {Ikeda}, N., {et~al.} 2009, in Astronomical
  Society of the Pacific Conference Series, Vol. 418, AKARI, a Light to
  Illuminate the Misty Universe, ed. T.~{Onaka}, G.~J. {White}, T.~{Nakagawa},
  \& I.~{Yamamura}, 3

\bibitem[{{Young} {et~al.}(2004){Young}, {J{\o}rgensen}, {Shirley},
  {Kauffmann}, {Huard}, {Lai}, {Lee}, {Crapsi}, {Bourke}, {Dullemond},
  {Brooke}, {Porras}, {Spiesman}, {Allen}, {Blake}, {Evans}, {Harvey},
  {Koerner}, {Mundy}, {Myers}, {Padgett}, {Sargent}, {Stapelfeldt}, {van
  Dishoeck}, {Bertoldi}, {Chapman}, {Cieza}, {DeVries}, {Ridge}, \&
  {Wahhaj}}]{Young2004}
{Young}, C.~H., {J{\o}rgensen}, J.~K., {Shirley}, Y.~L., {et~al.} 2004, \apjs,
  154, 396

\end{thebibliography}

\appendix
\section{ALMA observations centered at the APEX/LABOCA core detection coordinates.}
\begin{table*}
\caption{\label{ALMA} ALMA observation log}
\small
\centering
\begin{tabular}{llllll}
\hline\hline
ALMA pointing  & LABOCA Designation\tablefootmark{a} & RA (J2000)\tablefootmark{b} &DEC\tablefootmark{b} & continuum rms  & Detection?\\
               &                         & [h m s]  &  [$^\circ$ \,\arcmin \, \arcsec]           &[mJy/beam] &\\        
\hline
 
1 & LB30       &05:31:13.4   &+12:03:34.7  &0.13 &No\\
2 & LB31     & 05:31:15.3   &+12:03:38.7  &0.13 &YES\\
3 & LB29   &05:31:08.7   &+12:03:46.7 &0.13 &No\\
4 & LB22    &05:31:31.6   &+12:04:14.7  &0.13 &No\\
5  & LB24   &05:31:23.5   &+12:04:30.7  &0.13 &No\\
6 & LB23   &05:31:29.2   &+12.04.38.7  &0.13 &No\\
7 & LB21   &05:31:36.0   &+12:05:02.7  &0.13 &No\\ 
8 & LB28  &05:31:07.6   &+12:05:06.7  &0.13 &No\\
9 & LB26   &05:31:17.7   &+12:05:06.7  &0.13 &No\\
10 & LB25  &05:31:20.5   &+12:05:06.7  &0.14 &No\\
11 & LB27  &05:31:13.4   &+12:05:30.7  &0.13 &No\\ 
12  & LB20 &05:31:22.9   &+12:05:30.7  &0.12 &No\\
13   & LB19 &05:31:28.1   &+12:05:30.7  & 0.20 &YES\\
14    & LB18 &05:31:34.9   &+12:06:22.7  &0.12 &No\\
15    &LB17 & 05:31:09.8   &+12:06:42.7  &0.13 &No\\
16       & LB15        &05:31:27.3    &+12:07:26.7    &0.17  &No\\
17    &LB16      &05:31:30.0    &+12:07:06.7    &0.16  &No\\
18     & LB14 & 05:31:19.4    &+12:09:10.7    &0.25  &YES\\
19     & LB13 & 05:31:18.5    &+12:09:58.7    &0.17 &No\\
20       & LB12  &05:31:27.0    &+12:10:14.7    &0.17 &No\\
21     & LB11 & 05:31:10.4    &+12:10:38.7    &0.17 &No\\
22      & LB10 & 05:31:09.5    &+12:11:02.7    &0.23 &YES\\
23  &LB09 & 05:31:18.5    &+12:11:26.7    &0.17 &No\\
24 & LB08       &05:31:23.2    &+12:11:26.7    &0.26  &YES\\
25   & LB07 & 05:31:32.2    & +12:11:46.7    &0.17  &No\\
26     & LB06 & 05:31:30.8    &+12:12:46.7    &0.17 &No\\
27  & LB05        &05:31:29.5    &+12:14:30.7    &0.17  &No\\
28      & LB04 & 05:31:30.5    &+12:17:18.7    &0.17  &No\\
29   & LB03 & 05:31:29.2    &+12:17:22.7    &0.17  &No\\
30      & LB02 & 05:31:32.2    & +12:19:14.7   &0.17  &No\\

\hline
\hline                                                        
\end{tabular}                                                 
\tablefoot{ 
\tablefoottext{a}{From BGH16; } 
\tablefoottext{b}{Phase center coordinates} 
}

\end{table*}

\end{document}